\documentclass[aps,twocolumn,superscriptaddress,floatfix,longbibliography]{revtex4-2}
\usepackage{amsmath,amssymb,amsthm}
\usepackage{physics}
\usepackage{amsfonts}
\usepackage{mathrsfs}
\usepackage{graphicx}
\usepackage{tabularx}
\usepackage{enumerate}
\usepackage{dcolumn}
\usepackage{bm}
\usepackage{xcolor}
\usepackage[normalem]{ulem}
\usepackage[colorlinks,linkcolor=blue,citecolor=blue,urlcolor=blue]{hyperref}

\begin{document}
\title{Localization and mobility edges in non-Hermitian continuous quasiperiodic systems}

\author{Xiang-Ping Jiang}
\affiliation{Zhejiang Lab, Hangzhou 311121, China}

\author{Zhende Liu}
\address{Zhejiang Lab, Hangzhou 311121, China}

\author{Yayun Hu}
\email{yyhu@zhejianglab.edu.cn}
\affiliation{Zhejiang Lab, Hangzhou 311121, China}

\author{Lei Pan}
\email{panlei@nankai.edu.cn}
\affiliation{School of Physics, Nankai University, Tianjin 300071, China}

\date{\today}

\begin{abstract}
The mobility edge (ME) is a fundamental concept in the Anderson localized systems, which marks the energy separating extended and localized states. Although the ME and localization phenomena have been extensively studied in non-Hermitian (NH) quasiperiodic tight-binding models, they remain limited to NH continuum systems. Here, we investigate the ME and localization properties of a one-dimensional (1D) NH quasiperiodic continuous system, which is described by a Schr{\"o}dinger equation with an imaginary vector potential and an incommensurable one-site potential. We find that the ME locates in the real spectrum and falls between the localized and extended states. Additionally, we show that under the periodic boundary condition, the energy spectrum always exhibits an open curve representing high-energy extended electronic states characterized by a non-zero integer winding number. This complex spectrum topology is closely connected with the non-Hermitian skin effect (NHSE) observed under open boundary conditions, where the eigenstates of the bulk bands accumulate at the boundaries. Furthermore, we analyze the critical behavior of the localization transition and obtain critical potential amplitude accompanied by the universal critical exponent $\nu \simeq 1/3$. Our study provides valuable inspiration for exploring MEs and localization behaviors in NH quasiperiodic continuous systems.
\end{abstract}

\maketitle

\section{Introduction}
Anderson localization is a general phenomenon observed in quenched disorder systems, where the quantum interference can completely suppress the diffusion of single-particles~\cite{anderson1958absence,abrahams1979scaling,evers2008anderson}. This disorder-induced transition from metallic to insulating state typically manifests in dimensions exceeding two, although it can also be exhibited in one-dimensional quasiperiodic systems. The most well-known example of such a quasiperiodic system is the Aubry-Andr{\'e}-Harper (AAH) model~\cite{harper1955single,aubry1980analyticity}, whose localization transition point is independent of energy and is determined by the self-duality symmetry~\cite{gonccalves2022hidden,gonccalves2023critical,gonccalves2023renormalization,vu2023generic,wang2023exact}. However, it is vital to note that the self-duality symmetry in the AAH Hamiltonian can be broken under certain conditions. This symmetry breaking can lead to the emergence of an energy-dependent mobility edge (ME), which serves as a boundary between localized and extended states. The generalized AAH models with MEs have been extensively studied theoretically and experimentally~\cite{sarma1988mobility,boers2007mobility,biddle2009localization,ganeshan2015nearest,modugno2009exponential,biddle2010pre,luschen2018single,liu2018mobility,wang2020one,jiang2021mobility,zhang2022lyapunov,liu2022anomalous,wang2023two,wang2023engineering,zhou2023exact,qi2023multiple}.

In recent years, substantial advancements have been made in investigating non-Hermitian (NH) quantum systems~\cite{kunst2018biorthogonal,yao2018edge,gong2018topological}, discovering phenomena that lack a corresponding counterpart in Hermitian systems. Compared to Hermitian lattices, NH lattices exhibit topological non-trivial features even in one dimension and without symmetry~\cite{yokomizo2019non,kawabata2019symmetry,borgnia2020non,ashida2020non,bergholtz2021exceptional,ding2022non,okuma2023non,lin2023topological,yang2024homotopy}. This is attributable to the potential for forming closed loops in the complex plane, characterized by a nonzero spectrum winding number. The energy spectrum's point-gap topology under periodic boundary conditions (PBCs), which corresponds to a nonzero winding number, lies at the core of the non-Hermitian skin effect (NHSE)~\cite{yi2020non,zhang2020correspondence,okuma2020topological,li2020critical,zeng2022evolution,zeng2022real,liu2021non,longhi2022non,longhi2022self,yuce2022coexistence,mao2023non,wang2022non,mao2024liouvillian,xiao2024coexistence,hou2024dissolution}, i.e., the macroscopic condensation of bulk eigenstates at boundaries in a lattice under open boundary conditions (OBCs). The presence of the NHSE in NH tight-binding systems significantly influences the system's localization and transport properties~\cite{longhi2019probing,song2019non,xiao2020non,pan2020non,liu2020helical,xu2021dynamical,zhai2022nonequilibrium,li2022dynamic,liang2022dynamic,kawabata2023entanglement,li2024non}.
Furthermore, the NHSE, Anderson localization, and MEs interplay has been extensively investigated within the quasiperiodic tight-binding models~\cite{jiang2019interplay,longhi2019topological,longhi2019metal,liu2020non,liu2020generalized,zeng2020winding,zeng2020topological1,zeng2020topological2,liu2021localization,cai2022localization,jiang2021non,wu2021non,liu2021exacta,zhou2021non,jiang2021non,mu2022non,wang2022topological,weidemann2022topological,zhou2022topological,xu2022exact,qi2023localization,gandhi2023topological,zhu2023topological,jiang2024exact,acharya2024localization,padhan2024complete,wang2024non,li2024ring,wang2024exact}. This burgeoning domain of research has garnered significant attention in recent years. However, many previous studies on the localization transitions and the MEs have been mainly limited to the NH tight-binding models, and the NH continuous systems are still poorly unknown~\cite{longhi2021non,mochizuki2022band,yokomizo2022non,yuce2022non,yuce2022non,zeng2023gaussian,hu2023non}. Specifically, it remains unclear whether differences in localization transition and NHSE exist in the context of NH continuum quasiperiodic models, and whether the ME and its associated transition from real to complex spectrum can be extended to these systems. These problems motivate us to fill the gap and focus on the NH continuous quasiperiodic systems.

In this work, we study the localization transitions and the MEs in an NH continuum system with quasiperiodic potentials, in which non-Hermiticity is introduced by an imaginary vector potential $g$. We have discovered that the MEs lie within the real spectrum and confirm that they exist between the localized and extended states. Under PBC, we have found that the real energy spectrum is associated with the localized states, whereas the complex energy spectrum corresponds to the extended states. We also show that the energy spectrum is strongly sensitive to the boundary conditions and that under OBC one can observe the NHSE, with bulk modes squeezed toward the edge of the crystal. Correspondingly, the band structure displays a point-gap topology with a nonvanishing winding number under PBC. However, contrary to the tight-binding models, the energy spectrum of the highest energy band of the crystal describes an open (rather than a closed) curve in a complex energy plane, which is nevertheless still characterized by a nonvanishing integer winding number. As the imaginary vector potential $g$ increases, band merging is observed until a single energy band emerges, described by an open curve in a complex energy plane approaching the energy dispersion curve of the free-particle limit. Additionally, we numerically determine the critical quasiperiodic amplitude $V_c$ of localization transition from the scaling of the inverse participation ratio (IPR) and discover the universal critical exponent $\nu \simeq 1/3$, which is different from the tight-binding AAH model.

The remaining sections of our paper are organized as follows. In Sec. \ref{section2}, we give a quasiperiodic continuous Schrödinger equation with an imaginary vector potential. In Sec. \ref{section3}, we determine the MEs in the shallow potential regime and relate them to the real energy spectrum under PBC. In Sec. \ref{section4}, we examine NHSE and its associated point-gap topology in a continuous, incommensurate NH system. Further, we elaborate on the critical behavior of localization transitions in Sec. \ref{section5}. Lastly, we provide a comprehensive summary in Sec. \ref{section6}.

\section{The model Hamiltonian}\label{section2}
The starting point of our analysis is provided by the
Schr{\"o}dinger equation with an imaginary vector potential $g$ and a bichromatic
potential $V(x)$. The associateded Hamiltonian for a single-particle system is expressed as follows~\cite{khoudli2019critical}:
\begin{equation}\label{Hamiltonian}
\mathcal{H}_{g}\psi (x) = \frac{1}{2m}\left[-i\hbar \frac{\partial}{\partial x}+ig \right]^{2}\psi(x)+V(x) \psi(x),
\end{equation}
where $m$ and $\hbar$ are the particle mass and the reduced Planck constant, respectively. The bichromatic lattice potential is incommensurable with equal amplitudes and denoted as 
\begin{equation}\label{potential}
V(x) = \frac{V}{2}[\cos(2k_{1}x)+\cos(2k_{2}x+\phi)].
\end{equation}
Here, the quantity $V$ is the amplitudes characterized by two incommensurate periodic potentials with spatial periods $\pi/k_j$, and $k_2/k_1=r$ is an irrational number. The relative phase shift $\phi$ is irrelevant except for certain values. To avoid such cases, we adopt the values $r=(\sqrt{5}-1)/2$ and $\phi=1$, in our subsequent analysis. We solve the Schr{\"o}dinger eigen equation $\mathcal{H}_{g}\psi (x) \equiv E \psi (x)$ numerically by using exact diagonalization for the periodic boundary conditions (PBC): $\psi(0) =\psi(L)$. Here, $L$  denotes the system size and $E$ represents the eigenenergy. In this paper, we set $E_r=\hbar^2k_1^2/2m$ as the recoil energy, as none of the incommensurate potential $V(x) $'s periodic components dominate each other. 

To quantify the localized characteristics of the eigenstates in our NH continuum systems, we introduce the inverse participation ratio (IPR) \cite{evers2008anderson} as a quantitative descriptor. The IPR is defined as follows: 
\begin{equation}\label{IPR}
{\rm {IPR}}_n = \frac{\int dx\, \vert\psi^{L}_{n}(x)\psi^{R}_{n}(x)\vert^{2}}{\left[\int dx\, \vert\psi^{L}_{n}(x)\psi^{R}_{n}(x)\vert \right]^2},
\end{equation}
where $\psi^{L/R}_{n}(x)$ is the left/right wave function, and $n$ denotes the eigenenergy index. The IPR generally ranges from $0$ for eigenstates extending over all sites to $1$ for those localized around a specific site in the thermodynamic limit. Consequently, the IPR serves as a useful indicator for the localization transition.  However, it is noteworthy that IPR demonstrates a gradual transition in response to particle energy, and therefore, solely relying on IPR is insufficient to distinguish between extended and localized states on a large scale. To determine the ME precisely, we employ a rigorous approach involving the IPR's finite-size scaling and the fractal dimension (FD) calculation. The FD can be expressed mathematically as follows~\cite{deng2019one}:
\begin{equation}\label{FD}
\tau=-\lim_{L\rightarrow \infty}\frac{\mathrm{log(IPR)}}{{\mathrm{log}}(L)},
\end{equation}
where $L$ represents the size of the system. It has been established that for sufficiently large systems, the value of $\tau $ tends towards zero, indicating a localized state, while a value of $\tau $ close to one suggests an extended state. 

\begin{figure*}[t]
	\centering
	\includegraphics[width=0.32\textwidth]{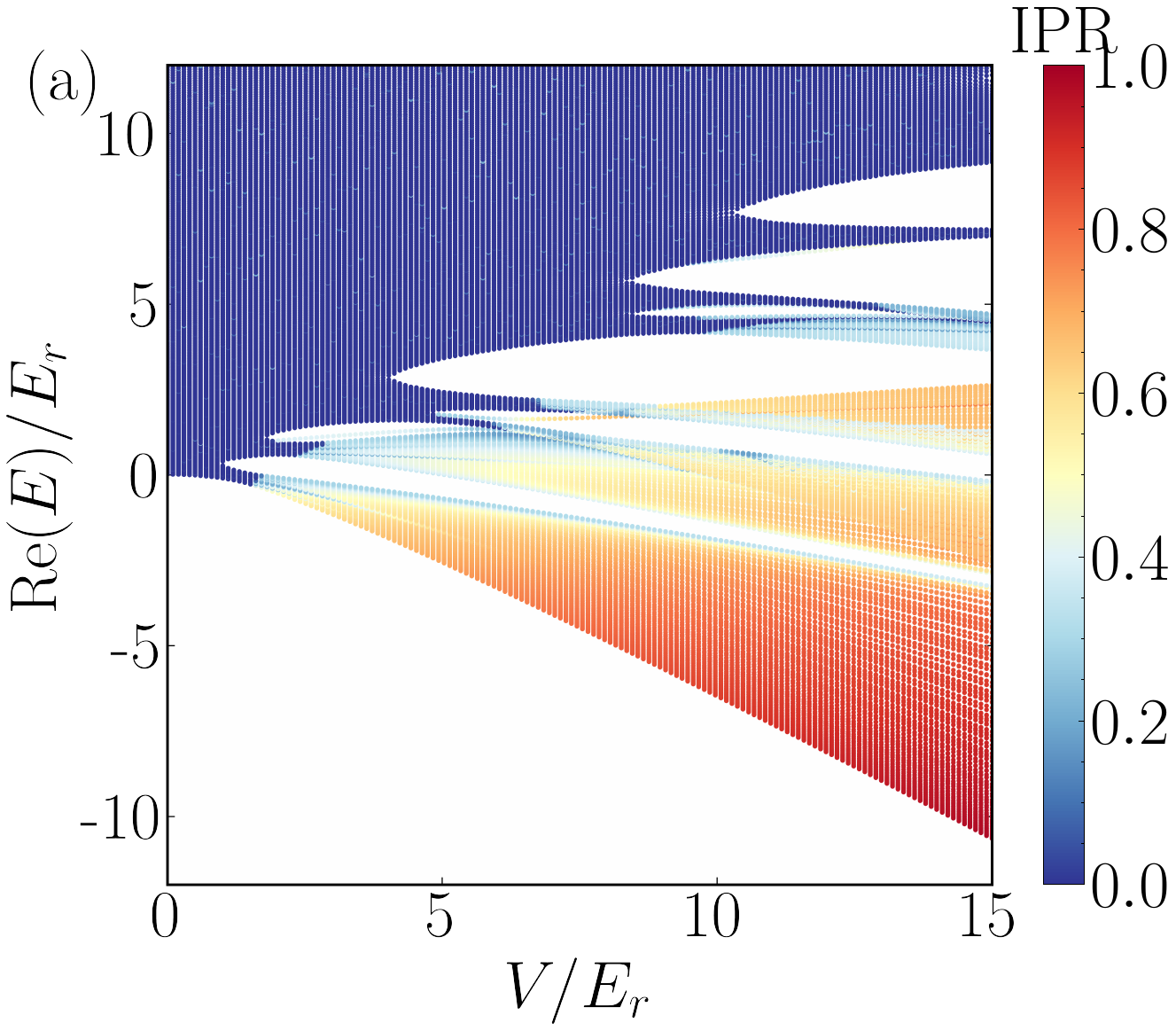}
    \includegraphics[width=0.32\textwidth]{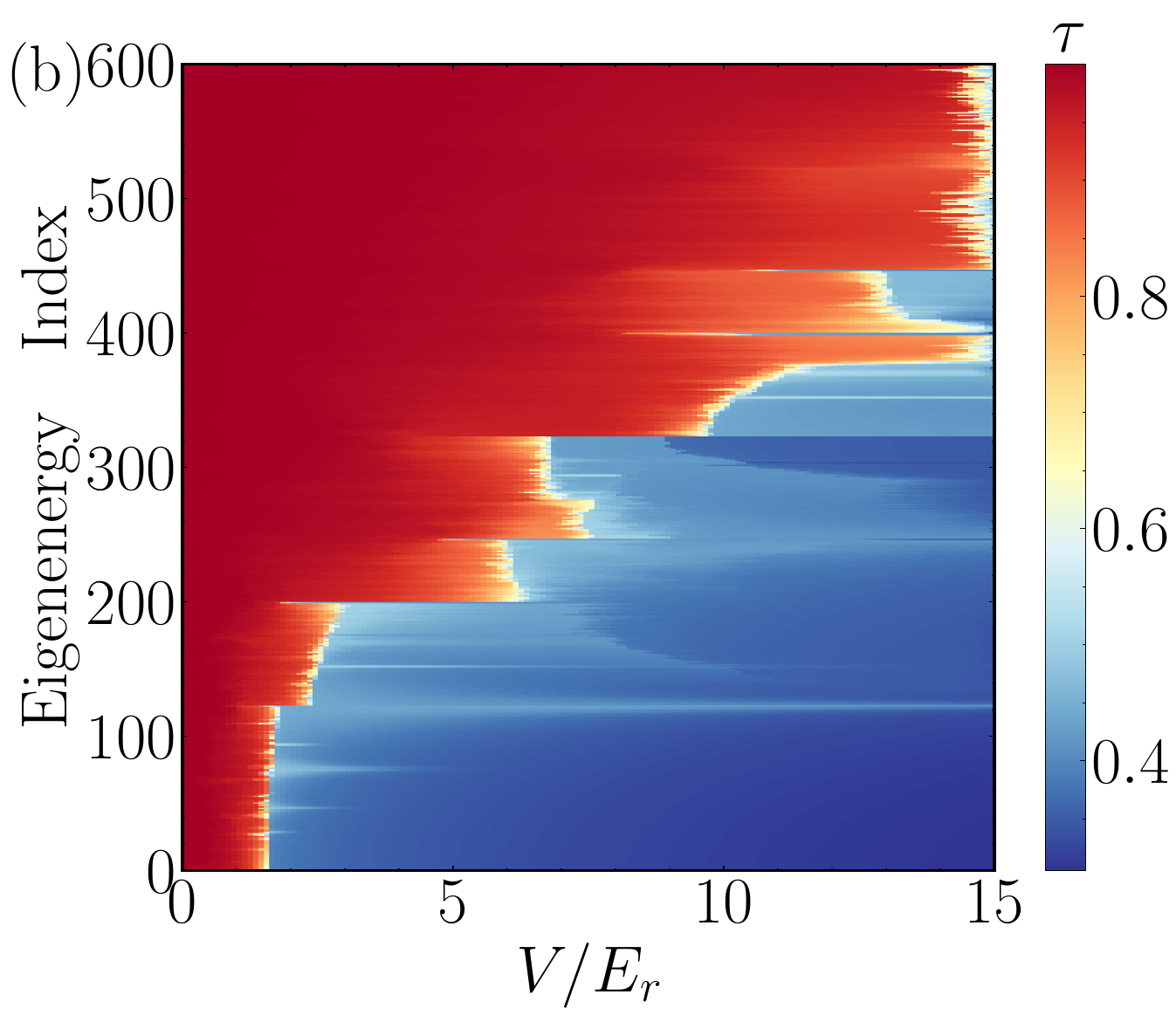}
    \includegraphics[width=0.33\textwidth]{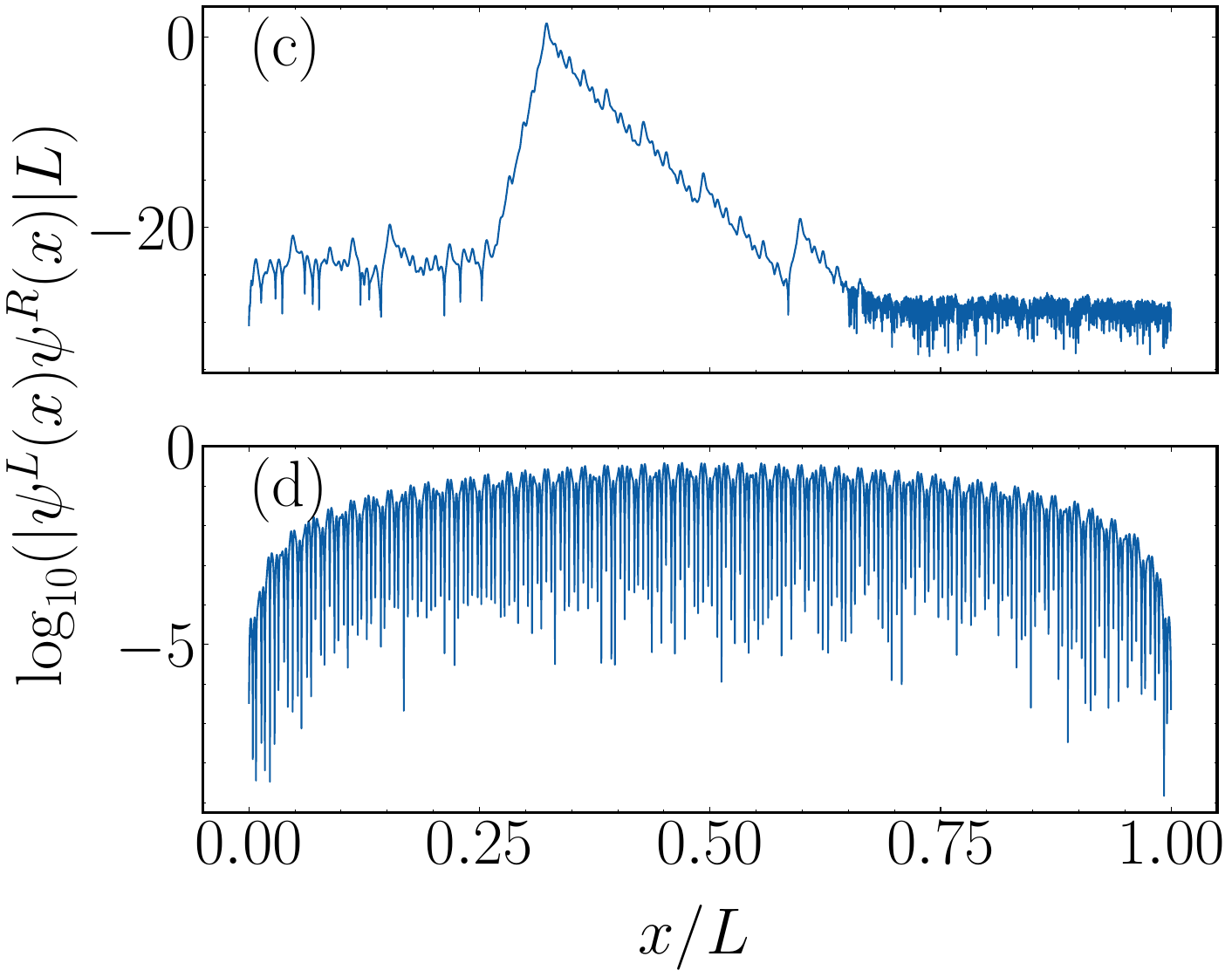}
	\caption{Localization transition and the MEs of the Hamiltonian (\ref{Hamiltonian}) with incommensurate potential $V(x)$. (a) The IPR versus the real energy spectrum ${\rm{Re}}(E)$ and the quasiperiodic amplitude $V$. Localized states within the system are characterized by large IPR values (red), while extended states are associated with nearly vanishingly small IPR values (blue). (b) The phase diagram of the energy spectrum presents a classification based on the ${\rm{Re}}(E)$ sorting index and the potential amplitude $V$. It is divided into two distinct regions: a high-energy extended region ($ \tau \simeq 0.9\pm0.1 $) and a low-energy localized region ($ \tau \simeq 0.4\pm0.1 $). (c) and (d) illustrate the density distribution of two typical eigenstates within the localized and extended regions, respectively. These states are chosen based on their energies, specifically below and above the ME at $V/E_r=2$. The system size $L=200a$ and the imaginary vector potential $g/E_r=2$ are used in our calculations.}
	\label{fig1}
\end{figure*}

\begin{figure}[b]
	\centering
	\includegraphics[width=0.48\textwidth]{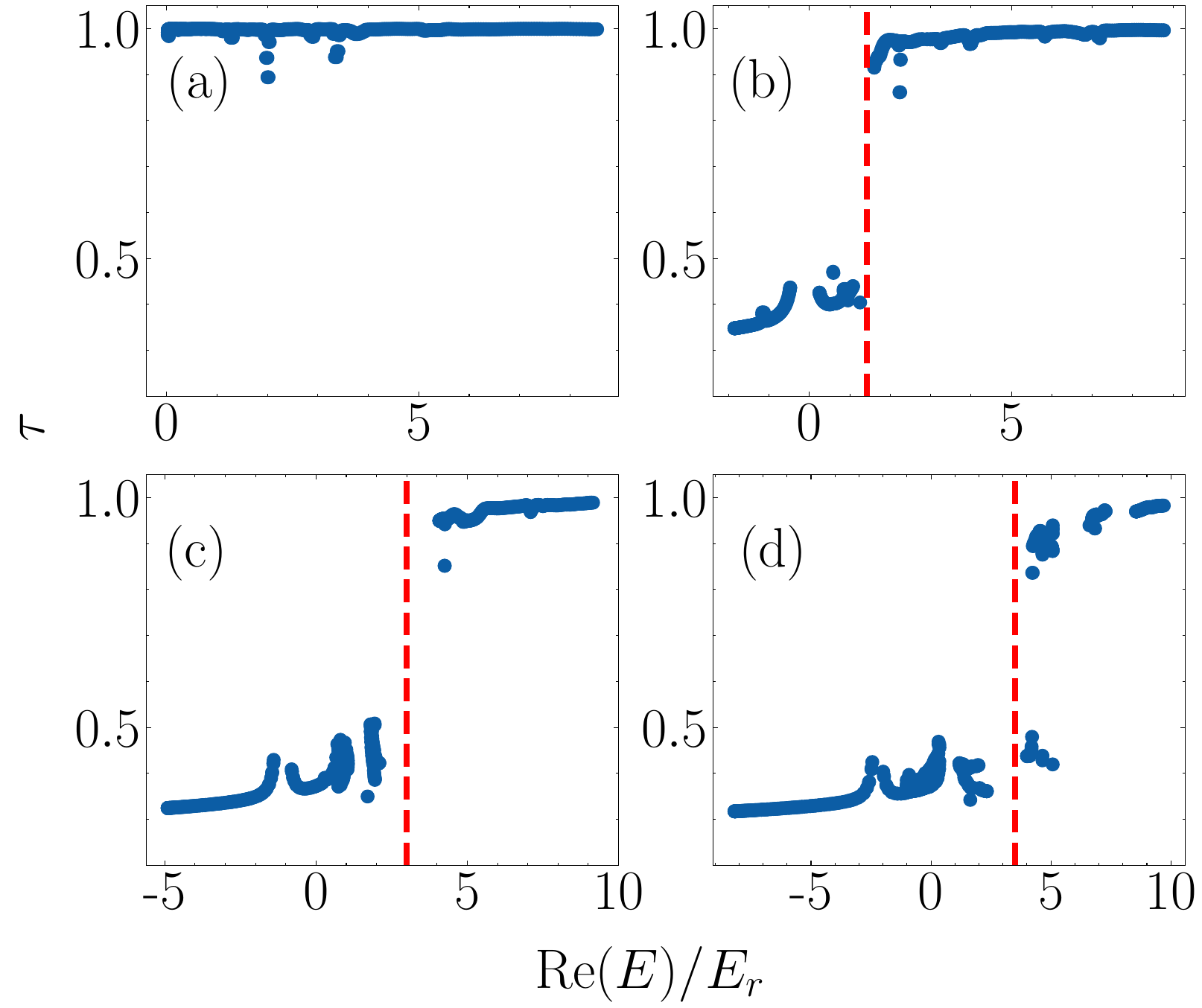}
	\caption{The fractal dimension $\tau $ of the real energy spectrum with the imaginary vector potential $g/E_r=2$ under PBC. (a) shows that all states are extended at $V/E_r=0$. (c)-(d) indicates the presence of MEs, and the corresponding parameters are $V/E_r=4.0,8.0,12.0$, respectively. The red dashed lines in the energy gap are the MEs.}
	\label{fig2}
\end{figure}

\section{Mobility edges in a non-Hermitian continuum system}\label{section3}
In this section, we computationally analyze the Hamiltonian given by equation (\ref{Hamiltonian}) using exact diagonalization, specifically focusing on the ME in the context of the NH quasiperiodic continuous system. The main numerical results are depicted in Fig.~\ref{fig1}. It is noteworthy to emphasize that the Hamiltonian represented in  (\ref{Hamiltonian}) cannot be mapped onto the tight-binding AAH model with non-reciprocal hopping terms., even when the potential amplitude $V \gg E_r$. As a result, the Hamiltonian under consideration has an energy-dependent ME. Figure \ref{fig1}(a) presents the $\rm {IPR}$ plotted against the real part of the energy spectrum, ${\rm{Re}}(E)$, and the potential amplitude $V$, for a substantial system size of $L=200a$, where $a=\pi/k_1$ represents the spatial periodicity of the first periodic potential and an imaginary vector potential $g=2E_r$ is employed. The numerical findings reveal the emergence of localization (characterized by high $\rm {IPR}$ values) at lower particle energies and elevated potential amplitudes, substantiating the presence of the $V$-dependent energy ME ${\rm{Re}}(E_c)$. This is further corroborated by the wave function behavior, transitioning from exponential localization at lower energies, as seen in Fig.~\ref{fig1}(c), to extended states at higher energies, as depicted in Fig.~\ref{fig1}(d). In the localized region, as depicted in Fig.~\ref{fig1}(c), it becomes evident that the eigenstates exhibit distinct characteristics depending on their proximity to a specific localized center, and this variation is influenced by the imaginary vector potential $g$. The imaginary vector potential induces the differing Lyapunov exponent of the left and right segments. 

Our analysis reveals that the observed ME presented in Fig.~\ref{fig1}(a) is in the energy gap for most values $V$ and the certain value $g/E_r=2$. Notably, all eigenstates manifest as extended states when $ V/E_r \lesssim 1.5$. However, a distinct ME signature emerges within the energy gap for some potential amplitudes $3.2 \lesssim V/E_r \lesssim 6.0$ and $7.0 \lesssim V/E_r \lesssim 10.5$. Conversely, the ME effect becomes more indistinct for other potential amplitudes, such as  $6.0 \lesssim V/E_r \lesssim7.0$ and $V/E_r \gtrsim  10.5$.

In our NH quasiperiodic continuous system, we find either $\tau \simeq 0.4\pm0.1$ (low-energy localized states) or $\tau \simeq 0.9\pm0.1$ (high-energy extended states). We observe a sharp change in the value of $\tau$; it shows the existence of an ME separating localized states and extended states. The ME ${\rm{Re}}(E_c)$ is then accurately determined as the real energy of the transition point between the two values. The numerical results are plotted in Fig.~\ref{fig2} for the system size $L=200a$ and the imaginary vector potential $g/E_r=2$. When $V/E_{r}= 0$, as shown in Fig.~\ref{fig2}(a), all states are extended. However, when increasing the incommensurable potential amplitude, as shown in Figs.~\ref{fig2}(b)-(d) there is a sharp transition for the value of $\tau$, and thus the MEs emerge. In all these cases, we find that the ME is always in an energy gap. In contrast, the ME is seen for some potential amplitudes [see Figs.~\ref{fig2}(b) and (c)]; it is more elusive for some other values [see Fig.~\ref{fig2}(d)]. In the latter case, it is a direct consequence of the energy spectrum's fractal behavior, as discussed in Ref.~\cite{khoudli2019critical}.

\section{Non-Hermitian skin effect, energy spectrum, and winding number in the NH continuous lattice}\label{section4}
In this section, let us examine the NHSE and the energy spectrum of the NH continuous incommensurate model (\ref{Hamiltonian}). Regarding the OBC, we can apply an imaginary gauge transformation, defined as $\psi (x)=\phi(x)e^{-gx}$, to the NH Hamiltonian (\ref{Hamiltonian}). This transformation reveals that the NH Hamiltonian transforms into a Hermitian Hamiltonian $\mathcal{H}_{g=0}$, and consequently, its energy spectrum becomes entirely real, composed of straight open curves. Consequently, the winding number of the energy spectrum under OBC is topologically trivial. However, the role of the imaginary vector potential goes beyond just changing the localization behaviors of eigenstate wave functions. Instead, it plays a pivotal role in the emergence of the NHSE through OBC. In fact, according to the imaginary gauge transformation, all extended Bloch eigenstates $\phi(x)$ in Hermitian limit exhibit exponentially localized behavior towards the boundary edge of the lattice, with a localization length scaling linearly to $\sim 1/|g|$. These results suggest that the emergence of the NHSE in finite lattices with OBC, induced by an imaginary vector potential, is a widely general feature. This holds not only in traditional tight-binding lattice models but also in continuous incommensurate systems.

\begin{figure}[t]
	\centering
	\includegraphics[width=0.50\textwidth]{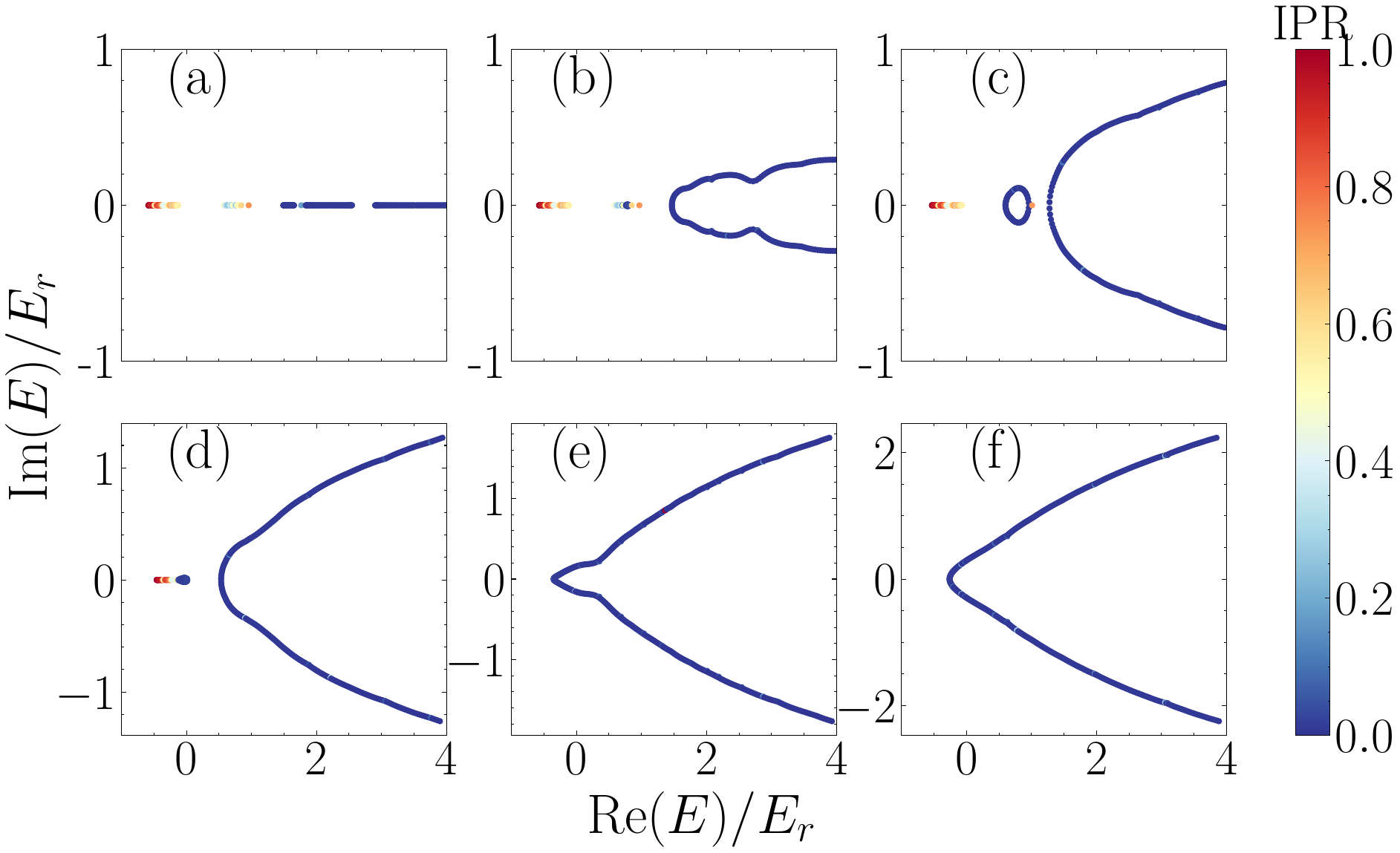}
	\caption{Energy spectrum under PBC of the NH continuous incommensurate potential (\ref{potential}). (a) $g/E_r=0$ (Hermitian case), (b) $g/E_r=1$, (c) $g/E_r=2$, (d) $g/E_r=3$, (e) $g/E_r=4$ and (f) $g/E_r=5$. The critical value $g_{c}/E_r\simeq 3.5$, above this value, the energy band will merge with the last closed curve and the open curve on the right. Here, the system size $L=200a$ and the NH incommensurate potential strength $V/E_r=2$.}
	\label{fig3}
\end{figure}

We subsequently examine the continuous NH Hamiltonian energy spectrum and non-trivial point-gap topology. Due to the nature of PBC, the imaginary gauge transformation applied under open boundary conditions (OBC) cannot be utilized to eliminate the imaginary field $g$. Consequently, the energy spectrum presents a set of complex curves in the energy planes, exhibiting non-trivial topology characterized by a non-vanishing spectral winding number. As per prior research, for any point-gap based energy $E_B$, a spectrum winding number can be defined for the PBC energy spectrum through the following relation~\cite{longhi2021non}:
\begin{equation}\label{winding}
\omega(E_B)= \frac{1}{2 \pi i }\sum  \int_{-\pi}^{\pi} dk \frac{d}{dk} \log [ E_{g}(k)-E_B ],
\end{equation}
where the summation is extended over various band intervals of the crystal. This relationship indicates how the energy spectrum winds around the point-gap energy as one varies certain parameters. We observe that $\omega(E_B) \neq 0$ whenever the based energy $E_B$ lies within any of the closed curves  $\mathcal{C}_c$ [as shown in Figs.~\ref{fig3}(c) or (d)], or on the rightmost side of the open curve $\mathcal{C}_o$, as shown in Figs.~\ref{fig3}(e)-(f). Even though curve $\mathcal{C}_o$ is open, the winding number $\omega$ remains quantized. This is intuitively attributed to the rapid closure of the curve at infinity, signified by $|{\rm{Im}}(E_{g}(k)) / {\rm{Re}}(E_{g}(k))| \rightarrow 0$ as one progresses along the curve $\mathcal{C}_o$ towards infinity. We numerically diagonalize NH Hamiltonian (\ref{Hamiltonian}) under PBC and get the results as shown in Fig.~\ref{fig3}. We can see that, for $g/E_r=0$ (Hermitian case) [Fig.~\ref{fig3}(a)], all energy spectra are real and low-energy eigenstates are localized, while high-energy eigenstates are extended. As the imaginary vector potential $g$ gradually increases, the energy spectra of the high-energy extended states are complex and form a parabolic open curve [Fig.~\ref{fig3}(b)]. When $g$ is moderately large, the energy spectrum in the middle part becomes a closed curve [Fig.~\ref{fig3}(c) and (d)]. When $g$ continues to increase and exceeds the critical value,
the last closed curve and the open curve on the right will merge and form a new open curve [Fig.~\ref{fig3}(e) and (f)]. Thus, as $g$ is increased and band merging occurs, a similar scenario is found, with $\omega(E_B) \neq 0$ when $E_B$ is chosen in the interior areas internal to the distinct bands [Fig.~\ref{fig3}(b) and (c)]. In particular, in the large $g$ limit $|\omega(E_B)|=1$ for any base energy $E_B$ in the interior of the parabolic curve [Fig.~\ref{fig3}(d)-(f)]. Once the winding number of the PBC energy spectrum is $\omega(E_B) \neq 0$, it indicates the existence of NHSE under OBC.

\begin{figure*}[t]
	\centering
	\includegraphics[width=0.95\textwidth]{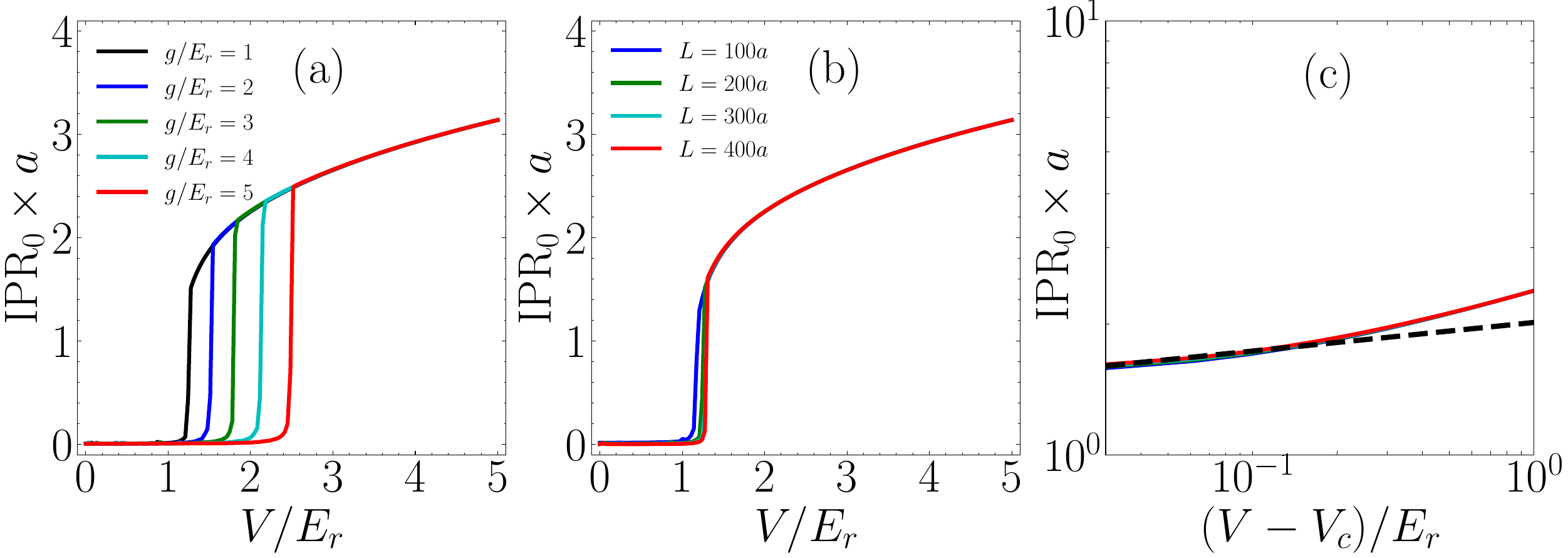}
	\caption{Critical behavior of localization transition. (a) Ground-state $\rm{IPR}_0$ versus the quasiperiodic amplitude $V/E_r$ for the different imaginary vector potentials strength $g/E_r$. (b) shows the $\rm{IPR}_0$ results for the different system sizes at a certain $g/E_r=1$. When the system size increases the $\rm{IPR}_0$ becomes critical in the larger system size. (c) Ground-state $\rm{IPR}_0$ versus $V-V_c$ in the log-log scale for our NH quasiperiodic continuous system. Here, the parameters are the same as (b).}
	\label{fig4}
\end{figure*}

\begin{figure}[!b]
	\centering
	\includegraphics[width=0.45\textwidth]{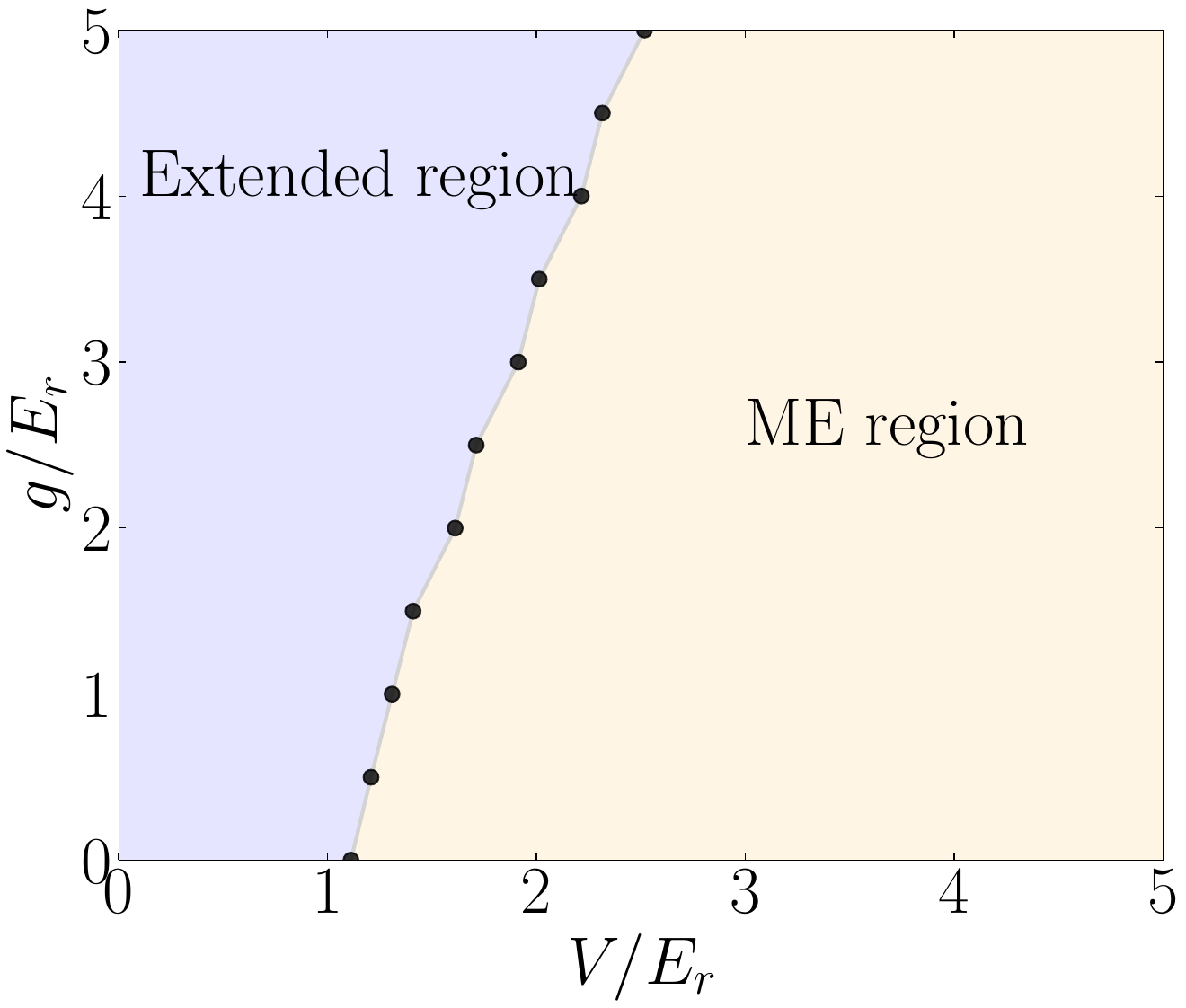}
	\caption{The phase diagram depicting the Hamiltonian (\ref{Hamiltonian}) as a function of the amplitude of the quasiperiodic potential $V$ and the imaginary vector potential $g$ within the context of PBC is presented. The gray region indicates the presence of MEs, and the remaining area is the extended region.}
	\label{fig5}
\end{figure}
\section{Critical behavior of the localization transition}\label{section5}
We now investigate the critical behavior of localization transition and construct the phase diagram within the $V-g$ parameter space for our NH quasiperiodic continuous system. As shown in Fig.~\ref{fig1}(a), a finite ME appears only for a potential amplitude $V$ larger than some critical value $V_c$ at a certain imaginary vector potential $g$. The occurrence of MEs in Hermitian incommensurate lattices has been documented in the literature~\cite{boers2007mobility,biddle2009localization,khoudli2019critical}. In Fig.~\ref{fig4}(a), we depict the IPR of the ground state ($\rm{IPR}_0$)  as a function of $V$ across various strengths of the imaginary vector potential $g/E_r$. Our findings reveal that the $\rm{IPR}_0$ exhibits a pronounced transition from close to $0$ to a finite value across all examined imaginary vector potential $g$ values. An increase in the imaginary vector potential $g$ is associated with a corresponding enhancement of the critical potential amplitude $V_c$. These numerical observations suggest that the imaginary vector potential within the quasiperiodic continuous system serves to inhibit localization. This phenomenon aligns with nonreciprocal hopping in NH quasiperiodic tight-binding models. Figure~\ref{fig4}(b) presents the IPR of the ground state ($\rm{IPR}_0$) as a function of  $V$ at the imaginary vector potential $g/E_r=1$, demonstrating a more pronounced transition from the extended to the localized phase with increasing system size, reaching a critical point in the thermodynamic limit. The scaling of the ground state IPR, $\rm{IPR}_0 \sim \mathcal{O}(1/L)$ in the extended region and $\rm{IPR}_0 \sim \mathcal{O}(1)$ in the localized region, enables the precise determination of the critical amplitude $V_c$ as the system size $L$ is augmented. By increasing the system size  $L$, we precisely determine the critical amplitude $V_c$ at the imaginary vector potential $g/E_r=1$, yielding
\begin{equation}\label{eq:criticalV0}
V_c/E_r \simeq 1.28\pm 0.02.
\end{equation}
Furthermore, this precise value of $V_c$ enables us to ascertain the critical exponent of the localization transition. Plotting $\rm{IPR}_0$ versus $V-V_c$ in log-log scale, we find a clear linear behavior for sufficiently large systems, consistent with the power-law scaling $\rm{IPR}_0 \propto (V-V_c)^{\nu}$: see Fig.~\ref{fig4}(c). By analyzing the slope of the fitting function, a critical exponent $\nu \simeq 0.326 \pm 0.004$ was determined. It is observed that for $V \gg V_c$, $E_r$, the behavior of the IPR changes. This is consistent with the exponent $1/4$ expected in the tight-binding limit, which differs from our NH continuous incommensurate system. It is worth noting that the critical exponent $\nu \simeq 1/3$ is universal and is the same as the Hermitian continuous quasiperiodic system. In Fig.~\ref{fig5}, we plot the phase diagram versus the imaginary vector potential $g$ and the quasiperiodic amplitude $V$. The gray area corresponds to the extended region where the ME is absent, and all low-energy states are extended states, while the remaining area corresponds to the ME region, the coexistence of extended and localized states.

\section{Conclusions}\label{section6}
In conclusion, we have explored the characteristics of the MEs, NHSE, and critical localization behavior of the 1D quasiperiodic continuous systems incorporating an imaginary vector potential, surpassing the conventional tight-binding approximation. Our results indicate that the MEs are always in the real spectrum gap and separate localized and extended states in the PBC of our NH quasiperiodic continuous system. We also find that this continuous system exhibits the NHSE under OBC, implying that the eigenstates are exponentially concentrated towards the lattice edges for any non-vanishing imaginary vector potential. The NHSE is characterized by a non-vanishing point-gap topological winding number within the PBC complex energy spectrum. Furthermore, we have found that the PBC energy spectrum in our quasiperiodic system consistently displays an open curve in the half-complex plane, which extends to infinity without bound. This behavior is consistent across all values of the imaginary vector potential and becomes more pronounced as the imaginary potential is increased to very large values. When the imaginary potential is sufficiently high,  this open curve characterizes the entire energy spectrum, resulting from a sequence of band-merging phenomena. Lastly, we have analyzed critical localization behavior and observed that the ground state $\rm{IPR}_0\propto (V-V_c)^{\nu}$ scaling and a universal critical exponent of approximately $\nu \simeq 1/3$ differs from that of the tight-binding AAH model.

The findings of this study offer an insightful exploration of MEs, NHSE, and critical localization behavior within NH continuous systems, transcending the limitations of traditional tight-binding models. The significance of these results is expected to encourage further theoretical and experimental investigations into NH continuous systems. For instance, it would be worthwhile to explore various types of NH terms and inhomogeneous imaginary vector potentials in continuous systems. Additionally, studying continuous two-dimensional NH systems could provide insight into the MEs and second-order NHSE observed within the tight-binding framework. Another promising avenue for future research is to extend these findings to continuous interacting models related to many-body localization.

\section*{Acknowledgments}
This work is supported by the China Postdoctoral Science Foundation (No.~2023M743267) and the National Natural Science Foundation of China (Grant No.~12304290 and No.~12204432). LP also acknowledges support from the Fundamental Research Funds for the Central Universities.

\bibliography{Localization}

\begin{thebibliography}{100}%
\makeatletter
\providecommand \@ifxundefined [1]{%
 \@ifx{#1\undefined}
}%
\providecommand \@ifnum [1]{%
 \ifnum #1\expandafter \@firstoftwo
 \else \expandafter \@secondoftwo
 \fi
}%
\providecommand \@ifx [1]{%
 \ifx #1\expandafter \@firstoftwo
 \else \expandafter \@secondoftwo
 \fi
}%
\providecommand \natexlab [1]{#1}%
\providecommand \enquote  [1]{``#1''}%
\providecommand \bibnamefont  [1]{#1}%
\providecommand \bibfnamefont [1]{#1}%
\providecommand \citenamefont [1]{#1}%
\providecommand \href@noop [0]{\@secondoftwo}%
\providecommand \href [0]{\begingroup \@sanitize@url \@href}%
\providecommand \@href[1]{\@@startlink{#1}\@@href}%
\providecommand \@@href[1]{\endgroup#1\@@endlink}%
\providecommand \@sanitize@url [0]{\catcode `\\12\catcode `\$12\catcode `\&12\catcode `\#12\catcode `\^12\catcode `\_12\catcode `\%12\relax}%
\providecommand \@@startlink[1]{}%
\providecommand \@@endlink[0]{}%
\providecommand \url  [0]{\begingroup\@sanitize@url \@url }%
\providecommand \@url [1]{\endgroup\@href {#1}{\urlprefix }}%
\providecommand \urlprefix  [0]{URL }%
\providecommand \Eprint [0]{\href }%
\providecommand \doibase [0]{https://doi.org/}%
\providecommand \selectlanguage [0]{\@gobble}%
\providecommand \bibinfo  [0]{\@secondoftwo}%
\providecommand \bibfield  [0]{\@secondoftwo}%
\providecommand \translation [1]{[#1]}%
\providecommand \BibitemOpen [0]{}%
\providecommand \bibitemStop [0]{}%
\providecommand \bibitemNoStop [0]{.\EOS\space}%
\providecommand \EOS [0]{\spacefactor3000\relax}%
\providecommand \BibitemShut  [1]{\csname bibitem#1\endcsname}%
\let\auto@bib@innerbib\@empty
\bibitem [{\citenamefont {Anderson}(1958)}]{anderson1958absence}%
  \BibitemOpen
  \bibfield  {author} {\bibinfo {author} {\bibfnamefont {P.~W.}\ \bibnamefont {Anderson}},\ }\bibfield  {title} {\bibinfo {title} {Absence of diffusion in certain random lattices},\ }\href {https://doi.org/10.1103/PhysRev.109.1492} {\bibfield  {journal} {\bibinfo  {journal} {Phys. Rev.}\ }\textbf {\bibinfo {volume} {109}},\ \bibinfo {pages} {1492} (\bibinfo {year} {1958})}\BibitemShut {NoStop}%
\bibitem [{\citenamefont {Abrahams}\ \emph {et~al.}(1979)\citenamefont {Abrahams}, \citenamefont {Anderson}, \citenamefont {Licciardello},\ and\ \citenamefont {Ramakrishnan}}]{abrahams1979scaling}%
  \BibitemOpen
  \bibfield  {author} {\bibinfo {author} {\bibfnamefont {E.}~\bibnamefont {Abrahams}}, \bibinfo {author} {\bibfnamefont {P.~W.}\ \bibnamefont {Anderson}}, \bibinfo {author} {\bibfnamefont {D.~C.}\ \bibnamefont {Licciardello}},\ and\ \bibinfo {author} {\bibfnamefont {T.~V.}\ \bibnamefont {Ramakrishnan}},\ }\bibfield  {title} {\bibinfo {title} {Scaling theory of localization: Absence of quantum diffusion in two dimensions},\ }\href {https://doi.org/10.1103/PhysRevLett.42.673} {\bibfield  {journal} {\bibinfo  {journal} {Phys. Rev. Lett.}\ }\textbf {\bibinfo {volume} {42}},\ \bibinfo {pages} {673} (\bibinfo {year} {1979})}\BibitemShut {NoStop}%
\bibitem [{\citenamefont {Evers}\ and\ \citenamefont {Mirlin}(2008)}]{evers2008anderson}%
  \BibitemOpen
  \bibfield  {author} {\bibinfo {author} {\bibfnamefont {F.}~\bibnamefont {Evers}}\ and\ \bibinfo {author} {\bibfnamefont {A.~D.}\ \bibnamefont {Mirlin}},\ }\bibfield  {title} {\bibinfo {title} {Anderson transitions},\ }\href {https://doi.org/10.1103/RevModPhys.80.1355} {\bibfield  {journal} {\bibinfo  {journal} {Rev. Mod. Phys.}\ }\textbf {\bibinfo {volume} {80}},\ \bibinfo {pages} {1355} (\bibinfo {year} {2008})}\BibitemShut {NoStop}%
\bibitem [{\citenamefont {Harper}(1955)}]{harper1955single}%
  \BibitemOpen
  \bibfield  {author} {\bibinfo {author} {\bibfnamefont {P.~G.}\ \bibnamefont {Harper}},\ }\bibfield  {title} {\bibinfo {title} {Single band motion of conduction electrons in a uniform magnetic field},\ }\href {https://iopscience.iop.org/article/10.1088/0370-1298/68/10/304/meta?casa_token=v97M2LoByEoAAAAA:YXBLaDkgSfiBBP3KCSunowkrVsGgu_ZaO5huAAHJ65x9car8DWrIEMCohg28PIRsRTYPmzmZbc0T} {\bibfield  {journal} {\bibinfo  {journal} {Proc. Phys. Soc., London Sect A}\ }\textbf {\bibinfo {volume} {68}},\ \bibinfo {pages} {874} (\bibinfo {year} {1955})}\BibitemShut {NoStop}%
\bibitem [{\citenamefont {Aubry}\ and\ \citenamefont {André}(1980)}]{aubry1980analyticity}%
  \BibitemOpen
  \bibfield  {author} {\bibinfo {author} {\bibfnamefont {S.}~\bibnamefont {Aubry}}\ and\ \bibinfo {author} {\bibfnamefont {G.}~\bibnamefont {André}},\ }\bibfield  {title} {\bibinfo {title} {Analyticity breaking and {Anderson} localization in incommensurate lattices},\ }\href {https://chaos.if.uj.edu.pl/~delande/Lectures/files/An.Is.Phys.Soc.pdf} {\bibfield  {journal} {\bibinfo  {journal} {Ann. Israel Phys. Soc}\ }\textbf {\bibinfo {volume} {3}},\ \bibinfo {pages} {133} (\bibinfo {year} {1980})}\BibitemShut {NoStop}%
\bibitem [{\citenamefont {Gonçalves}\ \emph {et~al.}(2022)\citenamefont {Gonçalves}, \citenamefont {Amorim}, \citenamefont {Castro},\ and\ \citenamefont {Ribeiro}}]{gonccalves2022hidden}%
  \BibitemOpen
  \bibfield  {author} {\bibinfo {author} {\bibfnamefont {M.}~\bibnamefont {Gonçalves}}, \bibinfo {author} {\bibfnamefont {B.}~\bibnamefont {Amorim}}, \bibinfo {author} {\bibfnamefont {E.}~\bibnamefont {Castro}},\ and\ \bibinfo {author} {\bibfnamefont {P.}~\bibnamefont {Ribeiro}},\ }\bibfield  {title} {\bibinfo {title} {Hidden dualities in {1D} quasiperiodic lattice models},\ }\href {https://doi.org/10.21468/SciPostPhys.13.3.046} {\bibfield  {journal} {\bibinfo  {journal} {SciPost Phys.}\ }\textbf {\bibinfo {volume} {13}},\ \bibinfo {pages} {046} (\bibinfo {year} {2022})}\BibitemShut {NoStop}%
\bibitem [{\citenamefont {Gonccalves}\ \emph {et~al.}(2023{\natexlab{a}})\citenamefont {Gonccalves}, \citenamefont {Amorim}, \citenamefont {Castro},\ and\ \citenamefont {Ribeiro}}]{gonccalves2023critical}%
  \BibitemOpen
  \bibfield  {author} {\bibinfo {author} {\bibfnamefont {M.}~\bibnamefont {Gonccalves}}, \bibinfo {author} {\bibfnamefont {B.}~\bibnamefont {Amorim}}, \bibinfo {author} {\bibfnamefont {E.~V.}\ \bibnamefont {Castro}},\ and\ \bibinfo {author} {\bibfnamefont {P.}~\bibnamefont {Ribeiro}},\ }\bibfield  {title} {\bibinfo {title} {Critical phase dualities in 1d exactly solvable quasiperiodic models},\ }\href {https://doi.org/10.1103/PhysRevLett.131.186303} {\bibfield  {journal} {\bibinfo  {journal} {Phys. Rev. Lett.}\ }\textbf {\bibinfo {volume} {131}},\ \bibinfo {pages} {186303} (\bibinfo {year} {2023}{\natexlab{a}})}\BibitemShut {NoStop}%
\bibitem [{\citenamefont {Gonccalves}\ \emph {et~al.}(2023{\natexlab{b}})\citenamefont {Gonccalves}, \citenamefont {Amorim}, \citenamefont {Castro},\ and\ \citenamefont {Ribeiro}}]{gonccalves2023renormalization}%
  \BibitemOpen
  \bibfield  {author} {\bibinfo {author} {\bibfnamefont {M.}~\bibnamefont {Gonccalves}}, \bibinfo {author} {\bibfnamefont {B.}~\bibnamefont {Amorim}}, \bibinfo {author} {\bibfnamefont {E.~V.}\ \bibnamefont {Castro}},\ and\ \bibinfo {author} {\bibfnamefont {P.}~\bibnamefont {Ribeiro}},\ }\bibfield  {title} {\bibinfo {title} {Renormalization group theory of one-dimensional quasiperiodic lattice models with commensurate approximants},\ }\href {https://doi.org/10.1103/PhysRevB.108.L100201} {\bibfield  {journal} {\bibinfo  {journal} {Phys. Rev. B}\ }\textbf {\bibinfo {volume} {108}},\ \bibinfo {pages} {L100201} (\bibinfo {year} {2023}{\natexlab{b}})}\BibitemShut {NoStop}%
\bibitem [{\citenamefont {Vu}\ and\ \citenamefont {Sarma}(2023)}]{vu2023generic}%
  \BibitemOpen
  \bibfield  {author} {\bibinfo {author} {\bibfnamefont {D.}~\bibnamefont {Vu}}\ and\ \bibinfo {author} {\bibfnamefont {S.~D.}\ \bibnamefont {Sarma}},\ }\bibfield  {title} {\bibinfo {title} {Generic mobility edges in several classes of duality-breaking one-dimensional quasiperiodic potentials},\ }\href {https://doi.org/10.1103/PhysRevB.107.224206} {\bibfield  {journal} {\bibinfo  {journal} {Phys. Rev. B}\ }\textbf {\bibinfo {volume} {107}},\ \bibinfo {pages} {224206} (\bibinfo {year} {2023})}\BibitemShut {NoStop}%
\bibitem [{\citenamefont {Wang}\ \emph {et~al.}(2023{\natexlab{a}})\citenamefont {Wang}, \citenamefont {Xia}, \citenamefont {You}, \citenamefont {Zheng},\ and\ \citenamefont {Zhou}}]{wang2023exact}%
  \BibitemOpen
  \bibfield  {author} {\bibinfo {author} {\bibfnamefont {Y.}~\bibnamefont {Wang}}, \bibinfo {author} {\bibfnamefont {X.}~\bibnamefont {Xia}}, \bibinfo {author} {\bibfnamefont {J.}~\bibnamefont {You}}, \bibinfo {author} {\bibfnamefont {Z.}~\bibnamefont {Zheng}},\ and\ \bibinfo {author} {\bibfnamefont {Q.}~\bibnamefont {Zhou}},\ }\bibfield  {title} {\bibinfo {title} {Exact mobility edges for {1D} quasiperiodic models},\ }\href {https://link.springer.com/article/10.1007/s00220-023-04695-9} {\bibfield  {journal} {\bibinfo  {journal} {Commun. Math. Phys.}\ } (\bibinfo {year} {2023}{\natexlab{a}})}\BibitemShut {NoStop}%
\bibitem [{\citenamefont {Sarma}\ \emph {et~al.}(1988)\citenamefont {Sarma}, \citenamefont {He},\ and\ \citenamefont {Xie}}]{sarma1988mobility}%
  \BibitemOpen
  \bibfield  {author} {\bibinfo {author} {\bibfnamefont {S.~D.}\ \bibnamefont {Sarma}}, \bibinfo {author} {\bibfnamefont {S.}~\bibnamefont {He}},\ and\ \bibinfo {author} {\bibfnamefont {X.~C.}\ \bibnamefont {Xie}},\ }\bibfield  {title} {\bibinfo {title} {Mobility edge in a model one-dimensional potential},\ }\href {https://doi.org/10.1103/PhysRevLett.61.2144} {\bibfield  {journal} {\bibinfo  {journal} {Phys. Rev. Lett.}\ }\textbf {\bibinfo {volume} {61}},\ \bibinfo {pages} {2144} (\bibinfo {year} {1988})}\BibitemShut {NoStop}%
\bibitem [{\citenamefont {Boers}\ \emph {et~al.}(2007)\citenamefont {Boers}, \citenamefont {Goedeke}, \citenamefont {Hinrichs},\ and\ \citenamefont {Holthaus}}]{boers2007mobility}%
  \BibitemOpen
  \bibfield  {author} {\bibinfo {author} {\bibfnamefont {D.~J.}\ \bibnamefont {Boers}}, \bibinfo {author} {\bibfnamefont {B.}~\bibnamefont {Goedeke}}, \bibinfo {author} {\bibfnamefont {D.}~\bibnamefont {Hinrichs}},\ and\ \bibinfo {author} {\bibfnamefont {M.}~\bibnamefont {Holthaus}},\ }\bibfield  {title} {\bibinfo {title} {Mobility edges in bichromatic optical lattices},\ }\href {https://doi.org/10.1103/PhysRevA.75.063404} {\bibfield  {journal} {\bibinfo  {journal} {Phys. Rev. A}\ }\textbf {\bibinfo {volume} {75}},\ \bibinfo {pages} {063404} (\bibinfo {year} {2007})}\BibitemShut {NoStop}%
\bibitem [{\citenamefont {Biddle}\ \emph {et~al.}(2009)\citenamefont {Biddle}, \citenamefont {Wang}, \citenamefont {Priour},\ and\ \citenamefont {Sarma}}]{biddle2009localization}%
  \BibitemOpen
  \bibfield  {author} {\bibinfo {author} {\bibfnamefont {J.}~\bibnamefont {Biddle}}, \bibinfo {author} {\bibfnamefont {B.}~\bibnamefont {Wang}}, \bibinfo {author} {\bibfnamefont {D.~J.}\ \bibnamefont {Priour}},\ and\ \bibinfo {author} {\bibfnamefont {S.~D.}\ \bibnamefont {Sarma}},\ }\bibfield  {title} {\bibinfo {title} {Localization in one-dimensional incommensurate lattices beyond the aubry-andr\'e model},\ }\href {https://doi.org/10.1103/PhysRevA.80.021603} {\bibfield  {journal} {\bibinfo  {journal} {Phys. Rev. A}\ }\textbf {\bibinfo {volume} {80}},\ \bibinfo {pages} {021603} (\bibinfo {year} {2009})}\BibitemShut {NoStop}%
\bibitem [{\citenamefont {Ganeshan}\ \emph {et~al.}(2015)\citenamefont {Ganeshan}, \citenamefont {Pixley},\ and\ \citenamefont {Sarma}}]{ganeshan2015nearest}%
  \BibitemOpen
  \bibfield  {author} {\bibinfo {author} {\bibfnamefont {S.}~\bibnamefont {Ganeshan}}, \bibinfo {author} {\bibfnamefont {J.~H.}\ \bibnamefont {Pixley}},\ and\ \bibinfo {author} {\bibfnamefont {S.~D.}\ \bibnamefont {Sarma}},\ }\bibfield  {title} {\bibinfo {title} {Nearest neighbor tight binding models with an exact mobility edge in one dimension},\ }\href {https://doi.org/10.1103/PhysRevLett.114.146601} {\bibfield  {journal} {\bibinfo  {journal} {Phys. Rev. Lett.}\ }\textbf {\bibinfo {volume} {114}},\ \bibinfo {pages} {146601} (\bibinfo {year} {2015})}\BibitemShut {NoStop}%
\bibitem [{\citenamefont {Modugno}(2009)}]{modugno2009exponential}%
  \BibitemOpen
  \bibfield  {author} {\bibinfo {author} {\bibfnamefont {M.}~\bibnamefont {Modugno}},\ }\bibfield  {title} {\bibinfo {title} {Exponential localization in one-dimensional quasi-periodic optical lattices},\ }\href {https://iopscience.iop.org/article/10.1088/1367-2630/11/3/033023} {\bibfield  {journal} {\bibinfo  {journal} {New J. Phys.}\ }\textbf {\bibinfo {volume} {11}},\ \bibinfo {pages} {033023} (\bibinfo {year} {2009})}\BibitemShut {NoStop}%
\bibitem [{\citenamefont {Biddle}\ and\ \citenamefont {Sarma}(2010)}]{biddle2010pre}%
  \BibitemOpen
  \bibfield  {author} {\bibinfo {author} {\bibfnamefont {J.}~\bibnamefont {Biddle}}\ and\ \bibinfo {author} {\bibfnamefont {S.~D.}\ \bibnamefont {Sarma}},\ }\bibfield  {title} {\bibinfo {title} {Predicted {Mobility Edges in One-Dimensional Incommensurate Optical Lattices: An Exactly Solvable Model of Anderson Localization}},\ }\href {https://link.aps.org/doi/10.1103/PhysRevLett.104.070601} {\bibfield  {journal} {\bibinfo  {journal} {Phys. Rev. Lett.}\ }\textbf {\bibinfo {volume} {104}},\ \bibinfo {pages} {070601} (\bibinfo {year} {2010})}\BibitemShut {NoStop}%
\bibitem [{\citenamefont {Lüschen}\ \emph {et~al.}(2018)\citenamefont {Lüschen}, \citenamefont {Scherg}, \citenamefont {Kohlert}, \citenamefont {Schreiber}, \citenamefont {Bordia}, \citenamefont {Li}, \citenamefont {Sarma},\ and\ \citenamefont {Bloch}}]{luschen2018single}%
  \BibitemOpen
  \bibfield  {author} {\bibinfo {author} {\bibfnamefont {H.~P.}\ \bibnamefont {Lüschen}}, \bibinfo {author} {\bibfnamefont {S.}~\bibnamefont {Scherg}}, \bibinfo {author} {\bibfnamefont {T.}~\bibnamefont {Kohlert}}, \bibinfo {author} {\bibfnamefont {M.}~\bibnamefont {Schreiber}}, \bibinfo {author} {\bibfnamefont {P.}~\bibnamefont {Bordia}}, \bibinfo {author} {\bibfnamefont {X.}~\bibnamefont {Li}}, \bibinfo {author} {\bibfnamefont {S.~D.}\ \bibnamefont {Sarma}},\ and\ \bibinfo {author} {\bibfnamefont {I.}~\bibnamefont {Bloch}},\ }\bibfield  {title} {\bibinfo {title} {Single-particle mobility edge in a one-dimensional quasiperiodic optical lattice},\ }\href {https://doi.org/10.1103/PhysRevLett.120.160404} {\bibfield  {journal} {\bibinfo  {journal} {Phys. Rev. Lett.}\ }\textbf {\bibinfo {volume} {120}},\ \bibinfo {pages} {160404} (\bibinfo {year} {2018})}\BibitemShut {NoStop}%
\bibitem [{\citenamefont {Liu}\ and\ \citenamefont {Guo}(2018)}]{liu2018mobility}%
  \BibitemOpen
  \bibfield  {author} {\bibinfo {author} {\bibfnamefont {T.}~\bibnamefont {Liu}}\ and\ \bibinfo {author} {\bibfnamefont {H.}~\bibnamefont {Guo}},\ }\bibfield  {title} {\bibinfo {title} {Mobility edges in off-diagonal disordered tight-binding models},\ }\href {https://doi.org/10.1103/PhysRevB.98.104201} {\bibfield  {journal} {\bibinfo  {journal} {Phys. Rev. B}\ }\textbf {\bibinfo {volume} {98}},\ \bibinfo {pages} {104201} (\bibinfo {year} {2018})}\BibitemShut {NoStop}%
\bibitem [{\citenamefont {Wang}\ \emph {et~al.}(2020)\citenamefont {Wang}, \citenamefont {Xia}, \citenamefont {Zhang}, \citenamefont {Yao}, \citenamefont {Chen}, \citenamefont {You}, \citenamefont {Zhou},\ and\ \citenamefont {Liu}}]{wang2020one}%
  \BibitemOpen
  \bibfield  {author} {\bibinfo {author} {\bibfnamefont {Y.}~\bibnamefont {Wang}}, \bibinfo {author} {\bibfnamefont {X.}~\bibnamefont {Xia}}, \bibinfo {author} {\bibfnamefont {L.}~\bibnamefont {Zhang}}, \bibinfo {author} {\bibfnamefont {H.}~\bibnamefont {Yao}}, \bibinfo {author} {\bibfnamefont {S.}~\bibnamefont {Chen}}, \bibinfo {author} {\bibfnamefont {J.}~\bibnamefont {You}}, \bibinfo {author} {\bibfnamefont {Q.}~\bibnamefont {Zhou}},\ and\ \bibinfo {author} {\bibfnamefont {X.-J.}\ \bibnamefont {Liu}},\ }\bibfield  {title} {\bibinfo {title} {{One-Dimensional Quasiperiodic Mosaic Lattice with Exact Mobility Edges}},\ }\href {https://doi.org/10.1103/PhysRevLett.125.196604} {\bibfield  {journal} {\bibinfo  {journal} {Phys. Rev. Lett.}\ }\textbf {\bibinfo {volume} {125}},\ \bibinfo {pages} {196604} (\bibinfo {year} {2020})}\BibitemShut {NoStop}%
\bibitem [{\citenamefont {Jiang}\ \emph {et~al.}(2021{\natexlab{a}})\citenamefont {Jiang}, \citenamefont {Qiao},\ and\ \citenamefont {Cao}}]{jiang2021mobility}%
  \BibitemOpen
  \bibfield  {author} {\bibinfo {author} {\bibfnamefont {X.-P.}\ \bibnamefont {Jiang}}, \bibinfo {author} {\bibfnamefont {Y.}~\bibnamefont {Qiao}},\ and\ \bibinfo {author} {\bibfnamefont {J.}~\bibnamefont {Cao}},\ }\bibfield  {title} {\bibinfo {title} {Mobility edges and reentrant localization in one-dimensional dimerized non-{Hermitian} quasiperiodic lattice},\ }\href {https://doi.org/10.1088/1674-1056/ac11e5} {\bibfield  {journal} {\bibinfo  {journal} {Chin. Phys. B}\ }\textbf {\bibinfo {volume} {30}},\ \bibinfo {pages} {097202} (\bibinfo {year} {2021}{\natexlab{a}})}\BibitemShut {NoStop}%
\bibitem [{\citenamefont {Zhang}\ and\ \citenamefont {Zhang}(2022)}]{zhang2022lyapunov}%
  \BibitemOpen
  \bibfield  {author} {\bibinfo {author} {\bibfnamefont {Y.-C.}\ \bibnamefont {Zhang}}\ and\ \bibinfo {author} {\bibfnamefont {Y.-Y.}\ \bibnamefont {Zhang}},\ }\bibfield  {title} {\bibinfo {title} {Lyapunov exponent, mobility edges, and critical region in the generalized {Aubry-Andr\'e} model with an unbounded quasiperiodic potential},\ }\href {https://doi.org/10.1103/PhysRevB.105.174206} {\bibfield  {journal} {\bibinfo  {journal} {Phys. Rev. B}\ }\textbf {\bibinfo {volume} {105}},\ \bibinfo {pages} {174206} (\bibinfo {year} {2022})}\BibitemShut {NoStop}%
\bibitem [{\citenamefont {Liu}\ \emph {et~al.}(2022)\citenamefont {Liu}, \citenamefont {Xia}, \citenamefont {Longhi},\ and\ \citenamefont {Sanchez-Palencia}}]{liu2022anomalous}%
  \BibitemOpen
  \bibfield  {author} {\bibinfo {author} {\bibfnamefont {T.}~\bibnamefont {Liu}}, \bibinfo {author} {\bibfnamefont {X.}~\bibnamefont {Xia}}, \bibinfo {author} {\bibfnamefont {S.}~\bibnamefont {Longhi}},\ and\ \bibinfo {author} {\bibfnamefont {L.}~\bibnamefont {Sanchez-Palencia}},\ }\bibfield  {title} {\bibinfo {title} {Anomalous mobility edges in one-dimensional quasiperiodic models},\ }\href {https://doi.org/10.21468/SciPostPhys.12.1.027} {\bibfield  {journal} {\bibinfo  {journal} {SciPost Phys.}\ }\textbf {\bibinfo {volume} {12}},\ \bibinfo {pages} {027} (\bibinfo {year} {2022})}\BibitemShut {NoStop}%
\bibitem [{\citenamefont {Wang}\ \emph {et~al.}(2023{\natexlab{b}})\citenamefont {Wang}, \citenamefont {Zhang}, \citenamefont {Wan}, \citenamefont {He},\ and\ \citenamefont {Wang}}]{wang2023two}%
  \BibitemOpen
  \bibfield  {author} {\bibinfo {author} {\bibfnamefont {Y.}~\bibnamefont {Wang}}, \bibinfo {author} {\bibfnamefont {L.}~\bibnamefont {Zhang}}, \bibinfo {author} {\bibfnamefont {Y.}~\bibnamefont {Wan}}, \bibinfo {author} {\bibfnamefont {Y.}~\bibnamefont {He}},\ and\ \bibinfo {author} {\bibfnamefont {Y.}~\bibnamefont {Wang}},\ }\bibfield  {title} {\bibinfo {title} {Two-dimensional vertex-decorated {Lieb} lattice with exact mobility edges and robust flat bands},\ }\href {https://doi.org/10.1103/PhysRevB.107.L140201} {\bibfield  {journal} {\bibinfo  {journal} {Phys. Rev. B}\ }\textbf {\bibinfo {volume} {107}},\ \bibinfo {pages} {L140201} (\bibinfo {year} {2023}{\natexlab{b}})}\BibitemShut {NoStop}%
\bibitem [{\citenamefont {Wang}\ \emph {et~al.}(2023{\natexlab{c}})\citenamefont {Wang}, \citenamefont {Zhang}, \citenamefont {Wang},\ and\ \citenamefont {Chen}}]{wang2023engineering}%
  \BibitemOpen
  \bibfield  {author} {\bibinfo {author} {\bibfnamefont {Z.}~\bibnamefont {Wang}}, \bibinfo {author} {\bibfnamefont {Y.}~\bibnamefont {Zhang}}, \bibinfo {author} {\bibfnamefont {L.}~\bibnamefont {Wang}},\ and\ \bibinfo {author} {\bibfnamefont {S.}~\bibnamefont {Chen}},\ }\bibfield  {title} {\bibinfo {title} {Engineering mobility in quasiperiodic lattices with exact mobility edges},\ }\href {https://doi.org/10.1103/PhysRevB.108.174202} {\bibfield  {journal} {\bibinfo  {journal} {Phys. Rev. B}\ }\textbf {\bibinfo {volume} {108}},\ \bibinfo {pages} {174202} (\bibinfo {year} {2023}{\natexlab{c}})}\BibitemShut {NoStop}%
\bibitem [{\citenamefont {Zhou}\ \emph {et~al.}(2023)\citenamefont {Zhou}, \citenamefont {Wang}, \citenamefont {Poon}, \citenamefont {Zhou},\ and\ \citenamefont {Liu}}]{zhou2023exact}%
  \BibitemOpen
  \bibfield  {author} {\bibinfo {author} {\bibfnamefont {X.-C.}\ \bibnamefont {Zhou}}, \bibinfo {author} {\bibfnamefont {Y.}~\bibnamefont {Wang}}, \bibinfo {author} {\bibfnamefont {T.-F.~J.}\ \bibnamefont {Poon}}, \bibinfo {author} {\bibfnamefont {Q.}~\bibnamefont {Zhou}},\ and\ \bibinfo {author} {\bibfnamefont {X.-J.}\ \bibnamefont {Liu}},\ }\bibfield  {title} {\bibinfo {title} {Exact new mobility edges between critical and localized states},\ }\href {https://doi.org/10.1103/PhysRevLett.131.176401} {\bibfield  {journal} {\bibinfo  {journal} {Phys. Rev. Lett.}\ }\textbf {\bibinfo {volume} {131}},\ \bibinfo {pages} {176401} (\bibinfo {year} {2023})}\BibitemShut {NoStop}%
\bibitem [{\citenamefont {Qi}\ \emph {et~al.}(2023{\natexlab{a}})\citenamefont {Qi}, \citenamefont {Cao},\ and\ \citenamefont {Jiang}}]{qi2023multiple}%
  \BibitemOpen
  \bibfield  {author} {\bibinfo {author} {\bibfnamefont {R.}~\bibnamefont {Qi}}, \bibinfo {author} {\bibfnamefont {J.}~\bibnamefont {Cao}},\ and\ \bibinfo {author} {\bibfnamefont {X.-P.}\ \bibnamefont {Jiang}},\ }\bibfield  {title} {\bibinfo {title} {Multiple localization transitions and novel quantum phases induced by a staggered on-site potential},\ }\href {https://doi.org/10.1103/PhysRevB.107.224201} {\bibfield  {journal} {\bibinfo  {journal} {Phys. Rev. B}\ }\textbf {\bibinfo {volume} {107}},\ \bibinfo {pages} {224201} (\bibinfo {year} {2023}{\natexlab{a}})}\BibitemShut {NoStop}%
\bibitem [{\citenamefont {Kunst}\ \emph {et~al.}(2018)\citenamefont {Kunst}, \citenamefont {Edvardsson}, \citenamefont {Budich},\ and\ \citenamefont {Bergholtz}}]{kunst2018biorthogonal}%
  \BibitemOpen
  \bibfield  {author} {\bibinfo {author} {\bibfnamefont {F.~K.}\ \bibnamefont {Kunst}}, \bibinfo {author} {\bibfnamefont {E.}~\bibnamefont {Edvardsson}}, \bibinfo {author} {\bibfnamefont {J.~C.}\ \bibnamefont {Budich}},\ and\ \bibinfo {author} {\bibfnamefont {E.~J.}\ \bibnamefont {Bergholtz}},\ }\bibfield  {title} {\bibinfo {title} {{Biorthogonal Bulk-Boundary Correspondence in Non-Hermitian Systems}},\ }\href {https://doi.org/10.1103/PhysRevLett.121.026808} {\bibfield  {journal} {\bibinfo  {journal} {Phys. Rev. Lett.}\ }\textbf {\bibinfo {volume} {121}},\ \bibinfo {pages} {026808} (\bibinfo {year} {2018})}\BibitemShut {NoStop}%
\bibitem [{\citenamefont {Yao}\ and\ \citenamefont {Wang}(2018)}]{yao2018edge}%
  \BibitemOpen
  \bibfield  {author} {\bibinfo {author} {\bibfnamefont {S.}~\bibnamefont {Yao}}\ and\ \bibinfo {author} {\bibfnamefont {Z.}~\bibnamefont {Wang}},\ }\bibfield  {title} {\bibinfo {title} {{Edge States and Topological Invariants of Non-Hermitian Systems}},\ }\href {https://doi.org/10.1103/PhysRevLett.121.086803} {\bibfield  {journal} {\bibinfo  {journal} {Phys. Rev. Lett.}\ }\textbf {\bibinfo {volume} {121}},\ \bibinfo {pages} {086803} (\bibinfo {year} {2018})}\BibitemShut {NoStop}%
\bibitem [{\citenamefont {Gong}\ \emph {et~al.}(2018)\citenamefont {Gong}, \citenamefont {Ashida}, \citenamefont {Kawabata}, \citenamefont {Takasan}, \citenamefont {Higashikawa},\ and\ \citenamefont {Ueda}}]{gong2018topological}%
  \BibitemOpen
  \bibfield  {author} {\bibinfo {author} {\bibfnamefont {Z.}~\bibnamefont {Gong}}, \bibinfo {author} {\bibfnamefont {Y.}~\bibnamefont {Ashida}}, \bibinfo {author} {\bibfnamefont {K.}~\bibnamefont {Kawabata}}, \bibinfo {author} {\bibfnamefont {K.}~\bibnamefont {Takasan}}, \bibinfo {author} {\bibfnamefont {S.}~\bibnamefont {Higashikawa}},\ and\ \bibinfo {author} {\bibfnamefont {M.}~\bibnamefont {Ueda}},\ }\bibfield  {title} {\bibinfo {title} {Topological phases of non-hermitian systems},\ }\href {https://doi.org/10.1103/PhysRevX.8.031079} {\bibfield  {journal} {\bibinfo  {journal} {Phys. Rev. X}\ }\textbf {\bibinfo {volume} {8}},\ \bibinfo {pages} {031079} (\bibinfo {year} {2018})}\BibitemShut {NoStop}%
\bibitem [{\citenamefont {Yokomizo}\ and\ \citenamefont {Murakami}(2019)}]{yokomizo2019non}%
  \BibitemOpen
  \bibfield  {author} {\bibinfo {author} {\bibfnamefont {K.}~\bibnamefont {Yokomizo}}\ and\ \bibinfo {author} {\bibfnamefont {S.}~\bibnamefont {Murakami}},\ }\bibfield  {title} {\bibinfo {title} {{Non-Bloch Band Theory of Non-Hermitian Systems}},\ }\href {https://doi.org/10.1103/PhysRevLett.123.066404} {\bibfield  {journal} {\bibinfo  {journal} {Phys. Rev. Lett.}\ }\textbf {\bibinfo {volume} {123}},\ \bibinfo {pages} {066404} (\bibinfo {year} {2019})}\BibitemShut {NoStop}%
\bibitem [{\citenamefont {Kawabata}\ \emph {et~al.}(2019)\citenamefont {Kawabata}, \citenamefont {Shiozaki}, \citenamefont {Ueda},\ and\ \citenamefont {Sato}}]{kawabata2019symmetry}%
  \BibitemOpen
  \bibfield  {author} {\bibinfo {author} {\bibfnamefont {K.}~\bibnamefont {Kawabata}}, \bibinfo {author} {\bibfnamefont {K.}~\bibnamefont {Shiozaki}}, \bibinfo {author} {\bibfnamefont {M.}~\bibnamefont {Ueda}},\ and\ \bibinfo {author} {\bibfnamefont {M.}~\bibnamefont {Sato}},\ }\bibfield  {title} {\bibinfo {title} {{Symmetry and Topology in Non-Hermitian Physics}},\ }\href {https://doi.org/10.1103/PhysRevX.9.041015} {\bibfield  {journal} {\bibinfo  {journal} {Phys. Rev. X}\ }\textbf {\bibinfo {volume} {9}},\ \bibinfo {pages} {041015} (\bibinfo {year} {2019})}\BibitemShut {NoStop}%
\bibitem [{\citenamefont {Borgnia}\ \emph {et~al.}(2020)\citenamefont {Borgnia}, \citenamefont {Kruchkov},\ and\ \citenamefont {Slager}}]{borgnia2020non}%
  \BibitemOpen
  \bibfield  {author} {\bibinfo {author} {\bibfnamefont {D.~S.}\ \bibnamefont {Borgnia}}, \bibinfo {author} {\bibfnamefont {A.~J.}\ \bibnamefont {Kruchkov}},\ and\ \bibinfo {author} {\bibfnamefont {R.-J.}\ \bibnamefont {Slager}},\ }\bibfield  {title} {\bibinfo {title} {{Non-Hermitian Boundary Modes and Topology}},\ }\href {https://doi.org/10.1103/PhysRevLett.124.056802} {\bibfield  {journal} {\bibinfo  {journal} {Phys. Rev. Lett.}\ }\textbf {\bibinfo {volume} {124}},\ \bibinfo {pages} {056802} (\bibinfo {year} {2020})}\BibitemShut {NoStop}%
\bibitem [{\citenamefont {Ashida}\ \emph {et~al.}(2020)\citenamefont {Ashida}, \citenamefont {Gong},\ and\ \citenamefont {Ueda}}]{ashida2020non}%
  \BibitemOpen
  \bibfield  {author} {\bibinfo {author} {\bibfnamefont {Y.}~\bibnamefont {Ashida}}, \bibinfo {author} {\bibfnamefont {Z.}~\bibnamefont {Gong}},\ and\ \bibinfo {author} {\bibfnamefont {M.}~\bibnamefont {Ueda}},\ }\bibfield  {title} {\bibinfo {title} {{Non-Hermitian Physics}},\ }\href {https://www.tandfonline.com/doi/abs/10.1080/00018732.2021.1876991} {\bibfield  {journal} {\bibinfo  {journal} {Adv. Phys.}\ }\textbf {\bibinfo {volume} {69}},\ \bibinfo {pages} {249} (\bibinfo {year} {2020})}\BibitemShut {NoStop}%
\bibitem [{\citenamefont {Bergholtz}\ \emph {et~al.}(2021)\citenamefont {Bergholtz}, \citenamefont {Budich},\ and\ \citenamefont {Kunst}}]{bergholtz2021exceptional}%
  \BibitemOpen
  \bibfield  {author} {\bibinfo {author} {\bibfnamefont {E.~J.}\ \bibnamefont {Bergholtz}}, \bibinfo {author} {\bibfnamefont {J.~C.}\ \bibnamefont {Budich}},\ and\ \bibinfo {author} {\bibfnamefont {F.~K.}\ \bibnamefont {Kunst}},\ }\bibfield  {title} {\bibinfo {title} {Exceptional topology of non-{Hermitian} systems},\ }\href {https://doi.org/10.1103/RevModPhys.93.015005} {\bibfield  {journal} {\bibinfo  {journal} {Rev. Mod. Phys.}\ }\textbf {\bibinfo {volume} {93}},\ \bibinfo {pages} {015005} (\bibinfo {year} {2021})}\BibitemShut {NoStop}%
\bibitem [{\citenamefont {Ding}\ \emph {et~al.}(2022)\citenamefont {Ding}, \citenamefont {Fang},\ and\ \citenamefont {Ma}}]{ding2022non}%
  \BibitemOpen
  \bibfield  {author} {\bibinfo {author} {\bibfnamefont {K.}~\bibnamefont {Ding}}, \bibinfo {author} {\bibfnamefont {C.}~\bibnamefont {Fang}},\ and\ \bibinfo {author} {\bibfnamefont {G.}~\bibnamefont {Ma}},\ }\bibfield  {title} {\bibinfo {title} {Non-{Hermitian} topology and exceptional-point geometries},\ }\href {https://www.nature.com/articles/s42254-022-00516-5} {\bibfield  {journal} {\bibinfo  {journal} {Nat. Rev. Phys.}\ }\textbf {\bibinfo {volume} {4}},\ \bibinfo {pages} {745} (\bibinfo {year} {2022})}\BibitemShut {NoStop}%
\bibitem [{\citenamefont {Okuma}\ and\ \citenamefont {Sato}(2023)}]{okuma2023non}%
  \BibitemOpen
  \bibfield  {author} {\bibinfo {author} {\bibfnamefont {N.}~\bibnamefont {Okuma}}\ and\ \bibinfo {author} {\bibfnamefont {M.}~\bibnamefont {Sato}},\ }\bibfield  {title} {\bibinfo {title} {Non-{Hermitian} topological phenomena: {A} review},\ }\href {https://www.annualreviews.org/content/journals/10.1146/annurev-conmatphys-040521-033133} {\bibfield  {journal} {\bibinfo  {journal} {Annual Review of Condensed Matter Physics}\ }\textbf {\bibinfo {volume} {14}},\ \bibinfo {pages} {83} (\bibinfo {year} {2023})}\BibitemShut {NoStop}%
\bibitem [{\citenamefont {Lin}\ \emph {et~al.}(2023)\citenamefont {Lin}, \citenamefont {Tai}, \citenamefont {Li},\ and\ \citenamefont {Lee}}]{lin2023topological}%
  \BibitemOpen
  \bibfield  {author} {\bibinfo {author} {\bibfnamefont {R.}~\bibnamefont {Lin}}, \bibinfo {author} {\bibfnamefont {T.}~\bibnamefont {Tai}}, \bibinfo {author} {\bibfnamefont {L.}~\bibnamefont {Li}},\ and\ \bibinfo {author} {\bibfnamefont {C.~H.}\ \bibnamefont {Lee}},\ }\bibfield  {title} {\bibinfo {title} {Topological non-{Hermitian} skin effect},\ }\href {https://link.springer.com/article/10.1007/s11467-023-1309-z} {\bibfield  {journal} {\bibinfo  {journal} {Frontiers of Physics}\ }\textbf {\bibinfo {volume} {18}},\ \bibinfo {pages} {53605} (\bibinfo {year} {2023})}\BibitemShut {NoStop}%
\bibitem [{\citenamefont {Yang}\ \emph {et~al.}(2024)\citenamefont {Yang}, \citenamefont {Li}, \citenamefont {König}, \citenamefont {Rødland}, \citenamefont {Stålhammar},\ and\ \citenamefont {Bergholtz}}]{yang2024homotopy}%
  \BibitemOpen
  \bibfield  {author} {\bibinfo {author} {\bibfnamefont {K.}~\bibnamefont {Yang}}, \bibinfo {author} {\bibfnamefont {Z.}~\bibnamefont {Li}}, \bibinfo {author} {\bibfnamefont {J.~L.~K.}\ \bibnamefont {König}}, \bibinfo {author} {\bibfnamefont {L.}~\bibnamefont {Rødland}}, \bibinfo {author} {\bibfnamefont {M.}~\bibnamefont {Stålhammar}},\ and\ \bibinfo {author} {\bibfnamefont {E.~J.}\ \bibnamefont {Bergholtz}},\ }\bibfield  {title} {\bibinfo {title} {Homotopy, symmetry, and non-{Hermitian} band topology},\ }\href {https://iopscience.iop.org/article/10.1088/1361-6633/ad4e64/meta} {\bibfield  {journal} {\bibinfo  {journal} {Rep. Prog. Phys.}\ }\textbf {\bibinfo {volume} {87}},\ \bibinfo {pages} {078002} (\bibinfo {year} {2024})}\BibitemShut {NoStop}%
\bibitem [{\citenamefont {Yi}\ and\ \citenamefont {Yang}(2020)}]{yi2020non}%
  \BibitemOpen
  \bibfield  {author} {\bibinfo {author} {\bibfnamefont {Y.}~\bibnamefont {Yi}}\ and\ \bibinfo {author} {\bibfnamefont {Z.}~\bibnamefont {Yang}},\ }\bibfield  {title} {\bibinfo {title} {{Non-Hermitian Skin Modes Induced by On-Site Dissipations and Chiral Tunneling Effect}},\ }\href {https://doi.org/10.1103/PhysRevLett.125.186802} {\bibfield  {journal} {\bibinfo  {journal} {Phys. Rev. Lett.}\ }\textbf {\bibinfo {volume} {125}},\ \bibinfo {pages} {186802} (\bibinfo {year} {2020})}\BibitemShut {NoStop}%
\bibitem [{\citenamefont {Zhang}\ \emph {et~al.}(2020)\citenamefont {Zhang}, \citenamefont {Yang},\ and\ \citenamefont {Fang}}]{zhang2020correspondence}%
  \BibitemOpen
  \bibfield  {author} {\bibinfo {author} {\bibfnamefont {K.}~\bibnamefont {Zhang}}, \bibinfo {author} {\bibfnamefont {Z.}~\bibnamefont {Yang}},\ and\ \bibinfo {author} {\bibfnamefont {C.}~\bibnamefont {Fang}},\ }\bibfield  {title} {\bibinfo {title} {{Correspondence between Winding Numbers and Skin Modes in Non-Hermitian Systems}},\ }\href {https://doi.org/10.1103/PhysRevLett.125.126402} {\bibfield  {journal} {\bibinfo  {journal} {Phys. Rev. Lett.}\ }\textbf {\bibinfo {volume} {125}},\ \bibinfo {pages} {126402} (\bibinfo {year} {2020})}\BibitemShut {NoStop}%
\bibitem [{\citenamefont {Okuma}\ \emph {et~al.}(2020)\citenamefont {Okuma}, \citenamefont {Kawabata}, \citenamefont {Shiozaki},\ and\ \citenamefont {Sato}}]{okuma2020topological}%
  \BibitemOpen
  \bibfield  {author} {\bibinfo {author} {\bibfnamefont {N.}~\bibnamefont {Okuma}}, \bibinfo {author} {\bibfnamefont {K.}~\bibnamefont {Kawabata}}, \bibinfo {author} {\bibfnamefont {K.}~\bibnamefont {Shiozaki}},\ and\ \bibinfo {author} {\bibfnamefont {M.}~\bibnamefont {Sato}},\ }\bibfield  {title} {\bibinfo {title} {{Topological Origin of Non-Hermitian Skin Effects}},\ }\href {https://doi.org/10.1103/PhysRevLett.124.086801} {\bibfield  {journal} {\bibinfo  {journal} {Phys. Rev. Lett.}\ }\textbf {\bibinfo {volume} {124}},\ \bibinfo {pages} {086801} (\bibinfo {year} {2020})}\BibitemShut {NoStop}%
\bibitem [{\citenamefont {Li}\ \emph {et~al.}(2020)\citenamefont {Li}, \citenamefont {Lee}, \citenamefont {Mu},\ and\ \citenamefont {Gong}}]{li2020critical}%
  \BibitemOpen
  \bibfield  {author} {\bibinfo {author} {\bibfnamefont {L.}~\bibnamefont {Li}}, \bibinfo {author} {\bibfnamefont {C.~H.}\ \bibnamefont {Lee}}, \bibinfo {author} {\bibfnamefont {S.}~\bibnamefont {Mu}},\ and\ \bibinfo {author} {\bibfnamefont {J.}~\bibnamefont {Gong}},\ }\bibfield  {title} {\bibinfo {title} {Critical non-{Hermitian} skin effect},\ }\href {https://www.nature.com/articles/s41467-020-18917-4} {\bibfield  {journal} {\bibinfo  {journal} {Nat. Commun.}\ }\textbf {\bibinfo {volume} {11}},\ \bibinfo {pages} {5491} (\bibinfo {year} {2020})}\BibitemShut {NoStop}%
\bibitem [{\citenamefont {Zeng}\ and\ \citenamefont {Lü}(2022{\natexlab{a}})}]{zeng2022evolution}%
  \BibitemOpen
  \bibfield  {author} {\bibinfo {author} {\bibfnamefont {Q.-B.}\ \bibnamefont {Zeng}}\ and\ \bibinfo {author} {\bibfnamefont {R.}~\bibnamefont {Lü}},\ }\bibfield  {title} {\bibinfo {title} {Evolution of spectral topology in one-dimensional long-range nonreciprocal lattices},\ }\href {https://doi.org/10.1103/PhysRevA.105.042211} {\bibfield  {journal} {\bibinfo  {journal} {Phys. Rev. A}\ }\textbf {\bibinfo {volume} {105}},\ \bibinfo {pages} {042211} (\bibinfo {year} {2022}{\natexlab{a}})}\BibitemShut {NoStop}%
\bibitem [{\citenamefont {Zeng}\ and\ \citenamefont {Lü}(2022{\natexlab{b}})}]{zeng2022real}%
  \BibitemOpen
  \bibfield  {author} {\bibinfo {author} {\bibfnamefont {Q.-B.}\ \bibnamefont {Zeng}}\ and\ \bibinfo {author} {\bibfnamefont {R.}~\bibnamefont {Lü}},\ }\bibfield  {title} {\bibinfo {title} {Real spectra and phase transition of skin effect in nonreciprocal systems},\ }\href {https://doi.org/10.1103/PhysRevB.105.245407} {\bibfield  {journal} {\bibinfo  {journal} {Phys. Rev. B}\ }\textbf {\bibinfo {volume} {105}},\ \bibinfo {pages} {245407} (\bibinfo {year} {2022}{\natexlab{b}})}\BibitemShut {NoStop}%
\bibitem [{\citenamefont {Liu}\ \emph {et~al.}(2021{\natexlab{a}})\citenamefont {Liu}, \citenamefont {Shao}, \citenamefont {Ma}, \citenamefont {Zhang}, \citenamefont {You}, \citenamefont {Wu}, \citenamefont {Xiang}, \citenamefont {Cui},\ and\ \citenamefont {Zhang}}]{liu2021non}%
  \BibitemOpen
  \bibfield  {author} {\bibinfo {author} {\bibfnamefont {S.}~\bibnamefont {Liu}}, \bibinfo {author} {\bibfnamefont {R.}~\bibnamefont {Shao}}, \bibinfo {author} {\bibfnamefont {S.}~\bibnamefont {Ma}}, \bibinfo {author} {\bibfnamefont {L.}~\bibnamefont {Zhang}}, \bibinfo {author} {\bibfnamefont {O.}~\bibnamefont {You}}, \bibinfo {author} {\bibfnamefont {H.}~\bibnamefont {Wu}}, \bibinfo {author} {\bibfnamefont {Y.~J.}\ \bibnamefont {Xiang}}, \bibinfo {author} {\bibfnamefont {T.~J.}\ \bibnamefont {Cui}},\ and\ \bibinfo {author} {\bibfnamefont {S.}~\bibnamefont {Zhang}},\ }\bibfield  {title} {\bibinfo {title} {Non-{Hermitian} skin effect in a non-{Hermitian} electrical circuit},\ }\href {https://spj.science.org/doi/full/10.34133/2021/5608038} {\bibfield  {journal} {\bibinfo  {journal} {Research}\ } (\bibinfo {year} {2021}{\natexlab{a}})}\BibitemShut {NoStop}%
\bibitem [{\citenamefont {Longhi}(2022{\natexlab{a}})}]{longhi2022non}%
  \BibitemOpen
  \bibfield  {author} {\bibinfo {author} {\bibfnamefont {S.}~\bibnamefont {Longhi}},\ }\bibfield  {title} {\bibinfo {title} {Non-{Hermitian} skin effect and self-acceleration},\ }\href {https://doi.org/10.1103/PhysRevB.105.245143} {\bibfield  {journal} {\bibinfo  {journal} {Phys. Rev. B}\ }\textbf {\bibinfo {volume} {105}},\ \bibinfo {pages} {245143} (\bibinfo {year} {2022}{\natexlab{a}})}\BibitemShut {NoStop}%
\bibitem [{\citenamefont {Longhi}(2022{\natexlab{b}})}]{longhi2022self}%
  \BibitemOpen
  \bibfield  {author} {\bibinfo {author} {\bibfnamefont {S.}~\bibnamefont {Longhi}},\ }\bibfield  {title} {\bibinfo {title} {{Self-Healing of Non-Hermitian Topological Skin Modes}},\ }\href {https://doi.org/10.1103/PhysRevLett.128.157601} {\bibfield  {journal} {\bibinfo  {journal} {Phys. Rev. Lett.}\ }\textbf {\bibinfo {volume} {128}},\ \bibinfo {pages} {157601} (\bibinfo {year} {2022}{\natexlab{b}})}\BibitemShut {NoStop}%
\bibitem [{\citenamefont {Yuce}\ and\ \citenamefont {Ramezani}(2022{\natexlab{a}})}]{yuce2022coexistence}%
  \BibitemOpen
  \bibfield  {author} {\bibinfo {author} {\bibfnamefont {C.}~\bibnamefont {Yuce}}\ and\ \bibinfo {author} {\bibfnamefont {H.}~\bibnamefont {Ramezani}},\ }\bibfield  {title} {\bibinfo {title} {Coexistence of extended and localized states in the one-dimensional non-hermitian anderson model},\ }\href {https://doi.org/10.1103/PhysRevB.106.024202} {\bibfield  {journal} {\bibinfo  {journal} {Phys. Rev. B}\ }\textbf {\bibinfo {volume} {106}},\ \bibinfo {pages} {024202} (\bibinfo {year} {2022}{\natexlab{a}})}\BibitemShut {NoStop}%
\bibitem [{\citenamefont {Mao}\ \emph {et~al.}(2023)\citenamefont {Mao}, \citenamefont {Hao},\ and\ \citenamefont {Pan}}]{mao2023non}%
  \BibitemOpen
  \bibfield  {author} {\bibinfo {author} {\bibfnamefont {L.}~\bibnamefont {Mao}}, \bibinfo {author} {\bibfnamefont {Y.}~\bibnamefont {Hao}},\ and\ \bibinfo {author} {\bibfnamefont {L.}~\bibnamefont {Pan}},\ }\bibfield  {title} {\bibinfo {title} {Non-hermitian skin effect in a one-dimensional interacting bose gas},\ }\href {https://doi.org/10.1103/PhysRevA.107.043315} {\bibfield  {journal} {\bibinfo  {journal} {Phys. Rev. A}\ }\textbf {\bibinfo {volume} {107}},\ \bibinfo {pages} {043315} (\bibinfo {year} {2023})}\BibitemShut {NoStop}%
\bibitem [{\citenamefont {Wang}\ \emph {et~al.}(2023{\natexlab{d}})\citenamefont {Wang}, \citenamefont {Suthar}, \citenamefont {Jen}, \citenamefont {Hsu},\ and\ \citenamefont {You}}]{wang2022non}%
  \BibitemOpen
  \bibfield  {author} {\bibinfo {author} {\bibfnamefont {Y.-C.}\ \bibnamefont {Wang}}, \bibinfo {author} {\bibfnamefont {K.}~\bibnamefont {Suthar}}, \bibinfo {author} {\bibfnamefont {H.~H.}\ \bibnamefont {Jen}}, \bibinfo {author} {\bibfnamefont {Y.-T.}\ \bibnamefont {Hsu}},\ and\ \bibinfo {author} {\bibfnamefont {J.-S.}\ \bibnamefont {You}},\ }\bibfield  {title} {\bibinfo {title} {Non-hermitian skin effects on thermal and many-body localized phases},\ }\href {https://doi.org/10.1103/PhysRevB.107.L220205} {\bibfield  {journal} {\bibinfo  {journal} {Phys. Rev. B}\ }\textbf {\bibinfo {volume} {107}},\ \bibinfo {pages} {L220205} (\bibinfo {year} {2023}{\natexlab{d}})}\BibitemShut {NoStop}%
\bibitem [{\citenamefont {Mao}\ \emph {et~al.}(2024)\citenamefont {Mao}, \citenamefont {Yang}, \citenamefont {Tao}, \citenamefont {Hu},\ and\ \citenamefont {Pan}}]{mao2024liouvillian}%
  \BibitemOpen
  \bibfield  {author} {\bibinfo {author} {\bibfnamefont {L.}~\bibnamefont {Mao}}, \bibinfo {author} {\bibfnamefont {X.}~\bibnamefont {Yang}}, \bibinfo {author} {\bibfnamefont {M.-J.}\ \bibnamefont {Tao}}, \bibinfo {author} {\bibfnamefont {H.}~\bibnamefont {Hu}},\ and\ \bibinfo {author} {\bibfnamefont {L.}~\bibnamefont {Pan}},\ }\bibfield  {title} {\bibinfo {title} {Liouvillian skin effect in a one-dimensional open many-body quantum system with generalized boundary conditions},\ }\href {https://doi.org/10.1103/PhysRevB.110.045440} {\bibfield  {journal} {\bibinfo  {journal} {Phys. Rev. B}\ }\textbf {\bibinfo {volume} {110}},\ \bibinfo {pages} {045440} (\bibinfo {year} {2024})}\BibitemShut {NoStop}%
\bibitem [{\citenamefont {Xiao}\ and\ \citenamefont {Zeng}(2024)}]{xiao2024coexistence}%
  \BibitemOpen
  \bibfield  {author} {\bibinfo {author} {\bibfnamefont {H.}~\bibnamefont {Xiao}}\ and\ \bibinfo {author} {\bibfnamefont {Q.-B.}\ \bibnamefont {Zeng}},\ }\bibfield  {title} {\bibinfo {title} {Coexistence of non-hermitian skin effect and extended states in one-dimensional nonreciprocal lattices},\ }\href {https://doi.org/10.1103/PhysRevB.110.024205} {\bibfield  {journal} {\bibinfo  {journal} {Phys. Rev. B}\ }\textbf {\bibinfo {volume} {110}},\ \bibinfo {pages} {024205} (\bibinfo {year} {2024})}\BibitemShut {NoStop}%
\bibitem [{\citenamefont {Hou}\ \emph {et~al.}(2024)\citenamefont {Hou}, \citenamefont {Xiao}, \citenamefont {Lü},\ and\ \citenamefont {Zeng}}]{hou2024dissolution}%
  \BibitemOpen
  \bibfield  {author} {\bibinfo {author} {\bibfnamefont {B.}~\bibnamefont {Hou}}, \bibinfo {author} {\bibfnamefont {H.}~\bibnamefont {Xiao}}, \bibinfo {author} {\bibfnamefont {R.}~\bibnamefont {Lü}},\ and\ \bibinfo {author} {\bibfnamefont {Q.-B.}\ \bibnamefont {Zeng}},\ }\bibfield  {title} {\bibinfo {title} {Dissolution of the non-hermitian skin effect in one-dimensional lattices with linearly varying nonreciprocal hopping},\ }\href {https://doi.org/10.1103/PhysRevB.109.094208} {\bibfield  {journal} {\bibinfo  {journal} {Phys. Rev. B}\ }\textbf {\bibinfo {volume} {109}},\ \bibinfo {pages} {094208} (\bibinfo {year} {2024})}\BibitemShut {NoStop}%
\bibitem [{\citenamefont {Longhi}(2019{\natexlab{a}})}]{longhi2019probing}%
  \BibitemOpen
  \bibfield  {author} {\bibinfo {author} {\bibfnamefont {S.}~\bibnamefont {Longhi}},\ }\bibfield  {title} {\bibinfo {title} {{Probing non-{Hermitian} skin effect and non-Bloch phase transitions}},\ }\href {https://doi.org/10.1103/PhysRevResearch.1.023013} {\bibfield  {journal} {\bibinfo  {journal} {Phys. Rev. Res.}\ }\textbf {\bibinfo {volume} {1}},\ \bibinfo {pages} {023013} (\bibinfo {year} {2019}{\natexlab{a}})}\BibitemShut {NoStop}%
\bibitem [{\citenamefont {Song}\ \emph {et~al.}(2019)\citenamefont {Song}, \citenamefont {Yao},\ and\ \citenamefont {Wang}}]{song2019non}%
  \BibitemOpen
  \bibfield  {author} {\bibinfo {author} {\bibfnamefont {F.}~\bibnamefont {Song}}, \bibinfo {author} {\bibfnamefont {S.}~\bibnamefont {Yao}},\ and\ \bibinfo {author} {\bibfnamefont {Z.}~\bibnamefont {Wang}},\ }\bibfield  {title} {\bibinfo {title} {{Non-Hermitian Skin Effect and Chiral Damping in Open Quantum Systems}},\ }\href {https://doi.org/10.1103/PhysRevLett.123.170401} {\bibfield  {journal} {\bibinfo  {journal} {Phys. Rev. Lett.}\ }\textbf {\bibinfo {volume} {123}},\ \bibinfo {pages} {170401} (\bibinfo {year} {2019})}\BibitemShut {NoStop}%
\bibitem [{\citenamefont {Xiao}\ \emph {et~al.}(2020)\citenamefont {Xiao}, \citenamefont {Deng}, \citenamefont {Wang}, \citenamefont {Zhu}, \citenamefont {Wang}, \citenamefont {Yi},\ and\ \citenamefont {Xue}}]{xiao2020non}%
  \BibitemOpen
  \bibfield  {author} {\bibinfo {author} {\bibfnamefont {L.}~\bibnamefont {Xiao}}, \bibinfo {author} {\bibfnamefont {T.}~\bibnamefont {Deng}}, \bibinfo {author} {\bibfnamefont {K.}~\bibnamefont {Wang}}, \bibinfo {author} {\bibfnamefont {G.}~\bibnamefont {Zhu}}, \bibinfo {author} {\bibfnamefont {Z.}~\bibnamefont {Wang}}, \bibinfo {author} {\bibfnamefont {W.}~\bibnamefont {Yi}},\ and\ \bibinfo {author} {\bibfnamefont {P.}~\bibnamefont {Xue}},\ }\bibfield  {title} {\bibinfo {title} {Non-hermitian bulk--boundary correspondence in quantum dynamics},\ }\href {https://www.nature.com/articles/s41567-020-0836-6} {\bibfield  {journal} {\bibinfo  {journal} {Nat. Phys.}\ }\textbf {\bibinfo {volume} {16}},\ \bibinfo {pages} {761} (\bibinfo {year} {2020})}\BibitemShut {NoStop}%
\bibitem [{\citenamefont {Pan}\ \emph {et~al.}(2020)\citenamefont {Pan}, \citenamefont {Chen}, \citenamefont {Chen},\ and\ \citenamefont {Zhai}}]{pan2020non}%
  \BibitemOpen
  \bibfield  {author} {\bibinfo {author} {\bibfnamefont {L.}~\bibnamefont {Pan}}, \bibinfo {author} {\bibfnamefont {X.}~\bibnamefont {Chen}}, \bibinfo {author} {\bibfnamefont {Y.}~\bibnamefont {Chen}},\ and\ \bibinfo {author} {\bibfnamefont {H.}~\bibnamefont {Zhai}},\ }\bibfield  {title} {\bibinfo {title} {Non-hermitian linear response theory},\ }\href {https://www.nature.com/articles/s41567-020-0889-6} {\bibfield  {journal} {\bibinfo  {journal} {Nat. Phys.}\ }\textbf {\bibinfo {volume} {16}},\ \bibinfo {pages} {767} (\bibinfo {year} {2020})}\BibitemShut {NoStop}%
\bibitem [{\citenamefont {Liu}\ \emph {et~al.}(2020{\natexlab{a}})\citenamefont {Liu}, \citenamefont {Zhang}, \citenamefont {Yang},\ and\ \citenamefont {Chen}}]{liu2020helical}%
  \BibitemOpen
  \bibfield  {author} {\bibinfo {author} {\bibfnamefont {C.-H.}\ \bibnamefont {Liu}}, \bibinfo {author} {\bibfnamefont {K.}~\bibnamefont {Zhang}}, \bibinfo {author} {\bibfnamefont {Z.}~\bibnamefont {Yang}},\ and\ \bibinfo {author} {\bibfnamefont {S.}~\bibnamefont {Chen}},\ }\bibfield  {title} {\bibinfo {title} {Helical damping and dynamical critical skin effect in open quantum systems},\ }\href {https://doi.org/10.1103/PhysRevResearch.2.043167} {\bibfield  {journal} {\bibinfo  {journal} {Phys. Rev. Res.}\ }\textbf {\bibinfo {volume} {2}},\ \bibinfo {pages} {043167} (\bibinfo {year} {2020}{\natexlab{a}})}\BibitemShut {NoStop}%
\bibitem [{\citenamefont {Xu}\ and\ \citenamefont {Chen}(2021)}]{xu2021dynamical}%
  \BibitemOpen
  \bibfield  {author} {\bibinfo {author} {\bibfnamefont {Z.}~\bibnamefont {Xu}}\ and\ \bibinfo {author} {\bibfnamefont {S.}~\bibnamefont {Chen}},\ }\bibfield  {title} {\bibinfo {title} {Dynamical evolution in a one-dimensional incommensurate lattice with $\mathcal{PT}$ symmetry},\ }\href {https://doi.org/10.1103/PhysRevA.103.043325} {\bibfield  {journal} {\bibinfo  {journal} {Phys. Rev. A}\ }\textbf {\bibinfo {volume} {103}},\ \bibinfo {pages} {043325} (\bibinfo {year} {2021})}\BibitemShut {NoStop}%
\bibitem [{\citenamefont {Zhai}\ \emph {et~al.}(2022)\citenamefont {Zhai}, \citenamefont {Huang},\ and\ \citenamefont {Yin}}]{zhai2022nonequilibrium}%
  \BibitemOpen
  \bibfield  {author} {\bibinfo {author} {\bibfnamefont {L.-J.}\ \bibnamefont {Zhai}}, \bibinfo {author} {\bibfnamefont {G.-Y.}\ \bibnamefont {Huang}},\ and\ \bibinfo {author} {\bibfnamefont {S.}~\bibnamefont {Yin}},\ }\bibfield  {title} {\bibinfo {title} {Nonequilibrium dynamics of the localization-delocalization transition in the non-{Hermitian Aubry-Andr\'e} model},\ }\href {https://doi.org/10.1103/PhysRevB.106.014204} {\bibfield  {journal} {\bibinfo  {journal} {Phys. Rev. B}\ }\textbf {\bibinfo {volume} {106}},\ \bibinfo {pages} {014204} (\bibinfo {year} {2022})}\BibitemShut {NoStop}%
\bibitem [{\citenamefont {Li}\ and\ \citenamefont {Wan}(2022)}]{li2022dynamic}%
  \BibitemOpen
  \bibfield  {author} {\bibinfo {author} {\bibfnamefont {H.}~\bibnamefont {Li}}\ and\ \bibinfo {author} {\bibfnamefont {S.}~\bibnamefont {Wan}},\ }\bibfield  {title} {\bibinfo {title} {Dynamic skin effects in non-{Hermitian} systems},\ }\href {https://doi.org/10.1103/PhysRevB.106.L241112} {\bibfield  {journal} {\bibinfo  {journal} {Phys. Rev. B}\ }\textbf {\bibinfo {volume} {106}},\ \bibinfo {pages} {L241112} (\bibinfo {year} {2022})}\BibitemShut {NoStop}%
\bibitem [{\citenamefont {Liang}\ \emph {et~al.}(2022)\citenamefont {Liang}, \citenamefont {Xie}, \citenamefont {Dong}, \citenamefont {Li}, \citenamefont {Li}, \citenamefont {Gadway}, \citenamefont {Yi},\ and\ \citenamefont {Yan}}]{liang2022dynamic}%
  \BibitemOpen
  \bibfield  {author} {\bibinfo {author} {\bibfnamefont {Q.}~\bibnamefont {Liang}}, \bibinfo {author} {\bibfnamefont {D.}~\bibnamefont {Xie}}, \bibinfo {author} {\bibfnamefont {Z.}~\bibnamefont {Dong}}, \bibinfo {author} {\bibfnamefont {H.}~\bibnamefont {Li}}, \bibinfo {author} {\bibfnamefont {H.}~\bibnamefont {Li}}, \bibinfo {author} {\bibfnamefont {B.}~\bibnamefont {Gadway}}, \bibinfo {author} {\bibfnamefont {W.}~\bibnamefont {Yi}},\ and\ \bibinfo {author} {\bibfnamefont {B.}~\bibnamefont {Yan}},\ }\bibfield  {title} {\bibinfo {title} {{Dynamic Signatures of Non-Hermitian Skin Effect and Topology in Ultracold Atoms}},\ }\href {https://doi.org/10.1103/PhysRevLett.129.070401} {\bibfield  {journal} {\bibinfo  {journal} {Phys. Rev. Lett.}\ }\textbf {\bibinfo {volume} {129}},\ \bibinfo {pages} {070401} (\bibinfo {year} {2022})}\BibitemShut {NoStop}%
\bibitem [{\citenamefont {Kawabata}\ \emph {et~al.}(2023)\citenamefont {Kawabata}, \citenamefont {Numasawa},\ and\ \citenamefont {Ryu}}]{kawabata2023entanglement}%
  \BibitemOpen
  \bibfield  {author} {\bibinfo {author} {\bibfnamefont {K.}~\bibnamefont {Kawabata}}, \bibinfo {author} {\bibfnamefont {T.}~\bibnamefont {Numasawa}},\ and\ \bibinfo {author} {\bibfnamefont {S.}~\bibnamefont {Ryu}},\ }\bibfield  {title} {\bibinfo {title} {Entanglement phase transition induced by the non-hermitian skin effect},\ }\href {https://doi.org/10.1103/PhysRevX.13.021007} {\bibfield  {journal} {\bibinfo  {journal} {Phys. Rev. X}\ }\textbf {\bibinfo {volume} {13}},\ \bibinfo {pages} {021007} (\bibinfo {year} {2023})}\BibitemShut {NoStop}%
\bibitem [{\citenamefont {Li}\ \emph {et~al.}(2024)\citenamefont {Li}, \citenamefont {Wang}, \citenamefont {Song},\ and\ \citenamefont {Wang}}]{li2024non}%
  \BibitemOpen
  \bibfield  {author} {\bibinfo {author} {\bibfnamefont {B.}~\bibnamefont {Li}}, \bibinfo {author} {\bibfnamefont {H.-R.}\ \bibnamefont {Wang}}, \bibinfo {author} {\bibfnamefont {F.}~\bibnamefont {Song}},\ and\ \bibinfo {author} {\bibfnamefont {Z.}~\bibnamefont {Wang}},\ }\bibfield  {title} {\bibinfo {title} {Non-bloch dynamics and topology in a classical nonequilibrium process},\ }\href {https://doi.org/10.1103/PhysRevB.109.L201121} {\bibfield  {journal} {\bibinfo  {journal} {Phys. Rev. B}\ }\textbf {\bibinfo {volume} {109}},\ \bibinfo {pages} {L201121} (\bibinfo {year} {2024})}\BibitemShut {NoStop}%
\bibitem [{\citenamefont {Jiang}\ \emph {et~al.}(2019)\citenamefont {Jiang}, \citenamefont {Lang}, \citenamefont {Yang}, \citenamefont {Zhu},\ and\ \citenamefont {Chen}}]{jiang2019interplay}%
  \BibitemOpen
  \bibfield  {author} {\bibinfo {author} {\bibfnamefont {H.}~\bibnamefont {Jiang}}, \bibinfo {author} {\bibfnamefont {L.-J.}\ \bibnamefont {Lang}}, \bibinfo {author} {\bibfnamefont {C.}~\bibnamefont {Yang}}, \bibinfo {author} {\bibfnamefont {S.-L.}\ \bibnamefont {Zhu}},\ and\ \bibinfo {author} {\bibfnamefont {S.}~\bibnamefont {Chen}},\ }\bibfield  {title} {\bibinfo {title} {Interplay of non-{Hermitian} skin effects and {Anderson} localization in nonreciprocal quasiperiodic lattices},\ }\href {https://doi.org/10.1103/PhysRevB.100.054301} {\bibfield  {journal} {\bibinfo  {journal} {Phys. Rev. B}\ }\textbf {\bibinfo {volume} {100}},\ \bibinfo {pages} {054301} (\bibinfo {year} {2019})}\BibitemShut {NoStop}%
\bibitem [{\citenamefont {Longhi}(2019{\natexlab{b}})}]{longhi2019topological}%
  \BibitemOpen
  \bibfield  {author} {\bibinfo {author} {\bibfnamefont {S.}~\bibnamefont {Longhi}},\ }\bibfield  {title} {\bibinfo {title} {{Topological Phase Transition in non-Hermitian Quasicrystals}},\ }\href {https://doi.org/10.1103/PhysRevLett.122.237601} {\bibfield  {journal} {\bibinfo  {journal} {Phys. Rev. Lett.}\ }\textbf {\bibinfo {volume} {122}},\ \bibinfo {pages} {237601} (\bibinfo {year} {2019}{\natexlab{b}})}\BibitemShut {NoStop}%
\bibitem [{\citenamefont {Longhi}(2019{\natexlab{c}})}]{longhi2019metal}%
  \BibitemOpen
  \bibfield  {author} {\bibinfo {author} {\bibfnamefont {S.}~\bibnamefont {Longhi}},\ }\bibfield  {title} {\bibinfo {title} {Metal-insulator phase transition in a non-{Hermitian} {Aubry-Andr\'{e}-Harper} model},\ }\href {https://doi.org/10.1103/PhysRevB.100.125157} {\bibfield  {journal} {\bibinfo  {journal} {Phys. Rev. B}\ }\textbf {\bibinfo {volume} {100}},\ \bibinfo {pages} {125157} (\bibinfo {year} {2019}{\natexlab{c}})}\BibitemShut {NoStop}%
\bibitem [{\citenamefont {Liu}\ \emph {et~al.}(2020{\natexlab{b}})\citenamefont {Liu}, \citenamefont {Jiang}, \citenamefont {Cao},\ and\ \citenamefont {Chen}}]{liu2020non}%
  \BibitemOpen
  \bibfield  {author} {\bibinfo {author} {\bibfnamefont {Y.}~\bibnamefont {Liu}}, \bibinfo {author} {\bibfnamefont {X.-P.}\ \bibnamefont {Jiang}}, \bibinfo {author} {\bibfnamefont {J.}~\bibnamefont {Cao}},\ and\ \bibinfo {author} {\bibfnamefont {S.}~\bibnamefont {Chen}},\ }\bibfield  {title} {\bibinfo {title} {Non-{Hermitian} mobility edges in one-dimensional quasicrystals with parity-time symmetry},\ }\href {https://doi.org/10.1103/PhysRevB.101.174205} {\bibfield  {journal} {\bibinfo  {journal} {Phys. Rev. B}\ }\textbf {\bibinfo {volume} {101}},\ \bibinfo {pages} {174205} (\bibinfo {year} {2020}{\natexlab{b}})}\BibitemShut {NoStop}%
\bibitem [{\citenamefont {Liu}\ \emph {et~al.}(2020{\natexlab{c}})\citenamefont {Liu}, \citenamefont {Guo}, \citenamefont {Pu},\ and\ \citenamefont {Longhi}}]{liu2020generalized}%
  \BibitemOpen
  \bibfield  {author} {\bibinfo {author} {\bibfnamefont {T.}~\bibnamefont {Liu}}, \bibinfo {author} {\bibfnamefont {H.}~\bibnamefont {Guo}}, \bibinfo {author} {\bibfnamefont {Y.}~\bibnamefont {Pu}},\ and\ \bibinfo {author} {\bibfnamefont {S.}~\bibnamefont {Longhi}},\ }\bibfield  {title} {\bibinfo {title} {Generalized {Aubry-Andr\'e} self-duality and mobility edges in non-{Hermitian} quasiperiodic lattices},\ }\href {https://doi.org/10.1103/PhysRevB.102.024205} {\bibfield  {journal} {\bibinfo  {journal} {Phys. Rev. B}\ }\textbf {\bibinfo {volume} {102}},\ \bibinfo {pages} {024205} (\bibinfo {year} {2020}{\natexlab{c}})}\BibitemShut {NoStop}%
\bibitem [{\citenamefont {Zeng}\ and\ \citenamefont {Xu}(2020)}]{zeng2020winding}%
  \BibitemOpen
  \bibfield  {author} {\bibinfo {author} {\bibfnamefont {Q.-B.}\ \bibnamefont {Zeng}}\ and\ \bibinfo {author} {\bibfnamefont {Y.}~\bibnamefont {Xu}},\ }\bibfield  {title} {\bibinfo {title} {Winding numbers and generalized mobility edges in non-{Hermitian} systems},\ }\href {https://doi.org/10.1103/PhysRevResearch.2.033052} {\bibfield  {journal} {\bibinfo  {journal} {Phys. Rev. Research}\ }\textbf {\bibinfo {volume} {2}},\ \bibinfo {pages} {033052} (\bibinfo {year} {2020})}\BibitemShut {NoStop}%
\bibitem [{\citenamefont {Zeng}\ \emph {et~al.}(2020{\natexlab{a}})\citenamefont {Zeng}, \citenamefont {Yang},\ and\ \citenamefont {Xu}}]{zeng2020topological1}%
  \BibitemOpen
  \bibfield  {author} {\bibinfo {author} {\bibfnamefont {Q.-B.}\ \bibnamefont {Zeng}}, \bibinfo {author} {\bibfnamefont {Y.-B.}\ \bibnamefont {Yang}},\ and\ \bibinfo {author} {\bibfnamefont {Y.}~\bibnamefont {Xu}},\ }\bibfield  {title} {\bibinfo {title} {Topological phases in non-{Hermitian} {Aubry-Andr\'e-Harper} models},\ }\href {https://doi.org/10.1103/PhysRevB.101.020201} {\bibfield  {journal} {\bibinfo  {journal} {Phys. Rev. B}\ }\textbf {\bibinfo {volume} {101}},\ \bibinfo {pages} {020201} (\bibinfo {year} {2020}{\natexlab{a}})}\BibitemShut {NoStop}%
\bibitem [{\citenamefont {Zeng}\ \emph {et~al.}(2020{\natexlab{b}})\citenamefont {Zeng}, \citenamefont {Yang},\ and\ \citenamefont {Lü}}]{zeng2020topological2}%
  \BibitemOpen
  \bibfield  {author} {\bibinfo {author} {\bibfnamefont {Q.-B.}\ \bibnamefont {Zeng}}, \bibinfo {author} {\bibfnamefont {Y.-B.}\ \bibnamefont {Yang}},\ and\ \bibinfo {author} {\bibfnamefont {R.}~\bibnamefont {Lü}},\ }\bibfield  {title} {\bibinfo {title} {Topological phases in one-dimensional nonreciprocal superlattices},\ }\href {https://doi.org/10.1103/PhysRevB.101.125418} {\bibfield  {journal} {\bibinfo  {journal} {Phys. Rev. B}\ }\textbf {\bibinfo {volume} {101}},\ \bibinfo {pages} {125418} (\bibinfo {year} {2020}{\natexlab{b}})}\BibitemShut {NoStop}%
\bibitem [{\citenamefont {Liu}\ \emph {et~al.}(2021{\natexlab{b}})\citenamefont {Liu}, \citenamefont {Zhou},\ and\ \citenamefont {Chen}}]{liu2021localization}%
  \BibitemOpen
  \bibfield  {author} {\bibinfo {author} {\bibfnamefont {Y.}~\bibnamefont {Liu}}, \bibinfo {author} {\bibfnamefont {Q.}~\bibnamefont {Zhou}},\ and\ \bibinfo {author} {\bibfnamefont {S.}~\bibnamefont {Chen}},\ }\bibfield  {title} {\bibinfo {title} {Localization transition, spectrum structure, and winding numbers for one-dimensional non-{Hermitian} quasicrystals},\ }\href {https://doi.org/10.1103/PhysRevB.104.024201} {\bibfield  {journal} {\bibinfo  {journal} {Phys. Rev. B}\ }\textbf {\bibinfo {volume} {104}},\ \bibinfo {pages} {024201} (\bibinfo {year} {2021}{\natexlab{b}})}\BibitemShut {NoStop}%
\bibitem [{\citenamefont {Cai}(2022)}]{cai2022localization}%
  \BibitemOpen
  \bibfield  {author} {\bibinfo {author} {\bibfnamefont {X.}~\bibnamefont {Cai}},\ }\bibfield  {title} {\bibinfo {title} {Localization transitions and winding numbers for {non-Hermitian Aubry-Andr\'e-Harper} models with off-diagonal modulations},\ }\href {https://doi.org/10.1103/PhysRevB.106.214207} {\bibfield  {journal} {\bibinfo  {journal} {Phys. Rev. B}\ }\textbf {\bibinfo {volume} {106}},\ \bibinfo {pages} {214207} (\bibinfo {year} {2022})}\BibitemShut {NoStop}%
\bibitem [{\citenamefont {Jiang}\ \emph {et~al.}(2021{\natexlab{b}})\citenamefont {Jiang}, \citenamefont {Qiao},\ and\ \citenamefont {Cao}}]{jiang2021non}%
  \BibitemOpen
  \bibfield  {author} {\bibinfo {author} {\bibfnamefont {X.-P.}\ \bibnamefont {Jiang}}, \bibinfo {author} {\bibfnamefont {Y.}~\bibnamefont {Qiao}},\ and\ \bibinfo {author} {\bibfnamefont {J.}~\bibnamefont {Cao}},\ }\bibfield  {title} {\bibinfo {title} {Non-{Hermitian Kitaev} chain with complex periodic and quasiperiodic potentials},\ }\href {https://doi.org/10.1088/1674-1056/abfa08} {\bibfield  {journal} {\bibinfo  {journal} {Chin. Phys. B.}\ }\textbf {\bibinfo {volume} {30}},\ \bibinfo {pages} {077101} (\bibinfo {year} {2021}{\natexlab{b}})}\BibitemShut {NoStop}%
\bibitem [{\citenamefont {Wu}\ \emph {et~al.}(2021)\citenamefont {Wu}, \citenamefont {Fan}, \citenamefont {Chen},\ and\ \citenamefont {Jia}}]{wu2021non}%
  \BibitemOpen
  \bibfield  {author} {\bibinfo {author} {\bibfnamefont {C.}~\bibnamefont {Wu}}, \bibinfo {author} {\bibfnamefont {J.}~\bibnamefont {Fan}}, \bibinfo {author} {\bibfnamefont {G.}~\bibnamefont {Chen}},\ and\ \bibinfo {author} {\bibfnamefont {S.}~\bibnamefont {Jia}},\ }\bibfield  {title} {\bibinfo {title} {Non-hermiticity-induced reentrant localization in a quasiperiodic lattice},\ }\href {https://doi.org/10.1088/1367-2630/ac430b} {\bibfield  {journal} {\bibinfo  {journal} {New J.Phys.}\ }\textbf {\bibinfo {volume} {23}},\ \bibinfo {pages} {123048} (\bibinfo {year} {2021})}\BibitemShut {NoStop}%
\bibitem [{\citenamefont {Liu}\ \emph {et~al.}(2021{\natexlab{c}})\citenamefont {Liu}, \citenamefont {Wang}, \citenamefont {Liu}, \citenamefont {Zhou},\ and\ \citenamefont {Chen}}]{liu2021exacta}%
  \BibitemOpen
  \bibfield  {author} {\bibinfo {author} {\bibfnamefont {Y.}~\bibnamefont {Liu}}, \bibinfo {author} {\bibfnamefont {Y.}~\bibnamefont {Wang}}, \bibinfo {author} {\bibfnamefont {X.-J.}\ \bibnamefont {Liu}}, \bibinfo {author} {\bibfnamefont {Q.}~\bibnamefont {Zhou}},\ and\ \bibinfo {author} {\bibfnamefont {S.}~\bibnamefont {Chen}},\ }\bibfield  {title} {\bibinfo {title} {Exact mobility edges, $\mathcal{PT}$-symmetry breaking, and skin effect in one-dimensional non-{Hermitian} quasicrystals},\ }\href {https://doi.org/10.1103/PhysRevB.103.014203} {\bibfield  {journal} {\bibinfo  {journal} {Phys. Rev. B}\ }\textbf {\bibinfo {volume} {103}},\ \bibinfo {pages} {014203} (\bibinfo {year} {2021}{\natexlab{c}})}\BibitemShut {NoStop}%
\bibitem [{\citenamefont {Zhou}\ and\ \citenamefont {Han}(2021)}]{zhou2021non}%
  \BibitemOpen
  \bibfield  {author} {\bibinfo {author} {\bibfnamefont {L.}~\bibnamefont {Zhou}}\ and\ \bibinfo {author} {\bibfnamefont {W.}~\bibnamefont {Han}},\ }\bibfield  {title} {\bibinfo {title} {Non-{Hermitian} quasicrystal in dimerized lattices},\ }\href {https://doi.org/10.1088/1674-1056/ac1efc} {\bibfield  {journal} {\bibinfo  {journal} {Chin. Phys. B.}\ }\textbf {\bibinfo {volume} {30}},\ \bibinfo {pages} {100308} (\bibinfo {year} {2021})}\BibitemShut {NoStop}%
\bibitem [{\citenamefont {Mu}\ \emph {et~al.}(2022)\citenamefont {Mu}, \citenamefont {Zhou}, \citenamefont {Li},\ and\ \citenamefont {Gong}}]{mu2022non}%
  \BibitemOpen
  \bibfield  {author} {\bibinfo {author} {\bibfnamefont {S.}~\bibnamefont {Mu}}, \bibinfo {author} {\bibfnamefont {L.}~\bibnamefont {Zhou}}, \bibinfo {author} {\bibfnamefont {L.}~\bibnamefont {Li}},\ and\ \bibinfo {author} {\bibfnamefont {J.}~\bibnamefont {Gong}},\ }\bibfield  {title} {\bibinfo {title} {Non-hermitian pseudo mobility edge in a coupled chain system},\ }\href {https://doi.org/10.1103/PhysRevB.105.205402} {\bibfield  {journal} {\bibinfo  {journal} {Phys. Rev. B}\ }\textbf {\bibinfo {volume} {105}},\ \bibinfo {pages} {205402} (\bibinfo {year} {2022})}\BibitemShut {NoStop}%
\bibitem [{\citenamefont {Wang}\ \emph {et~al.}(2022)\citenamefont {Wang}, \citenamefont {Xu}, \citenamefont {Li}, \citenamefont {Xu},\ and\ \citenamefont {Wang}}]{wang2022topological}%
  \BibitemOpen
  \bibfield  {author} {\bibinfo {author} {\bibfnamefont {Z.-H.}\ \bibnamefont {Wang}}, \bibinfo {author} {\bibfnamefont {F.}~\bibnamefont {Xu}}, \bibinfo {author} {\bibfnamefont {L.}~\bibnamefont {Li}}, \bibinfo {author} {\bibfnamefont {D.-H.}\ \bibnamefont {Xu}},\ and\ \bibinfo {author} {\bibfnamefont {B.}~\bibnamefont {Wang}},\ }\bibfield  {title} {\bibinfo {title} {Topological superconductors and exact mobility edges in non-{Hermitian} quasicrystals},\ }\href {https://doi.org/10.1103/PhysRevB.105.024514} {\bibfield  {journal} {\bibinfo  {journal} {Phys. Rev. B}\ }\textbf {\bibinfo {volume} {105}},\ \bibinfo {pages} {024514} (\bibinfo {year} {2022})}\BibitemShut {NoStop}%
\bibitem [{\citenamefont {Weidemann}\ \emph {et~al.}(2022)\citenamefont {Weidemann}, \citenamefont {Kremer}, \citenamefont {Longhi},\ and\ \citenamefont {Szameit}}]{weidemann2022topological}%
  \BibitemOpen
  \bibfield  {author} {\bibinfo {author} {\bibfnamefont {S.}~\bibnamefont {Weidemann}}, \bibinfo {author} {\bibfnamefont {M.}~\bibnamefont {Kremer}}, \bibinfo {author} {\bibfnamefont {S.}~\bibnamefont {Longhi}},\ and\ \bibinfo {author} {\bibfnamefont {A.}~\bibnamefont {Szameit}},\ }\bibfield  {title} {\bibinfo {title} {Topological triple phase transition in non-hermitian floquet quasicrystals},\ }\href {https://www.nature.com/articles/s41586-021-04253-0} {\bibfield  {journal} {\bibinfo  {journal} {Nature}\ }\textbf {\bibinfo {volume} {601}},\ \bibinfo {pages} {354} (\bibinfo {year} {2022})}\BibitemShut {NoStop}%
\bibitem [{\citenamefont {Wang}\ \emph {et~al.}(2021)\citenamefont {Wang}, \citenamefont {Xu}, \citenamefont {Li}, \citenamefont {Xu},\ and\ \citenamefont {Wang}}]{zhou2022topological}%
  \BibitemOpen
  \bibfield  {author} {\bibinfo {author} {\bibfnamefont {Z.-H.}\ \bibnamefont {Wang}}, \bibinfo {author} {\bibfnamefont {F.}~\bibnamefont {Xu}}, \bibinfo {author} {\bibfnamefont {L.}~\bibnamefont {Li}}, \bibinfo {author} {\bibfnamefont {D.-H.}\ \bibnamefont {Xu}},\ and\ \bibinfo {author} {\bibfnamefont {B.}~\bibnamefont {Wang}},\ }\bibfield  {title} {\bibinfo {title} {Unconventional real-complex spectral transition and majorana zero modes in nonreciprocal quasicrystals},\ }\href {https://doi.org/10.1103/PhysRevB.104.174501} {\bibfield  {journal} {\bibinfo  {journal} {Phys. Rev. B}\ }\textbf {\bibinfo {volume} {104}},\ \bibinfo {pages} {174501} (\bibinfo {year} {2021})}\BibitemShut {NoStop}%
\bibitem [{\citenamefont {Xu}\ \emph {et~al.}(2022)\citenamefont {Xu}, \citenamefont {Xia},\ and\ \citenamefont {Chen}}]{xu2022exact}%
  \BibitemOpen
  \bibfield  {author} {\bibinfo {author} {\bibfnamefont {Z.-H.}\ \bibnamefont {Xu}}, \bibinfo {author} {\bibfnamefont {X.}~\bibnamefont {Xia}},\ and\ \bibinfo {author} {\bibfnamefont {S.}~\bibnamefont {Chen}},\ }\bibfield  {title} {\bibinfo {title} {Exact mobility edges and topological phase transition in two-dimensional non-hermitian quasicrystals},\ }\href {https://doi.org/10.1007/s11433-021-1802-4} {\bibfield  {journal} {\bibinfo  {journal} {Sci. China Phys. Mech. Astron.}\ }\textbf {\bibinfo {volume} {65}},\ \bibinfo {pages} {227211} (\bibinfo {year} {2022})}\BibitemShut {NoStop}%
\bibitem [{\citenamefont {Qi}\ \emph {et~al.}(2023{\natexlab{b}})\citenamefont {Qi}, \citenamefont {Cao},\ and\ \citenamefont {Jiang}}]{qi2023localization}%
  \BibitemOpen
  \bibfield  {author} {\bibinfo {author} {\bibfnamefont {R.}~\bibnamefont {Qi}}, \bibinfo {author} {\bibfnamefont {J.}~\bibnamefont {Cao}},\ and\ \bibinfo {author} {\bibfnamefont {X.-P.}\ \bibnamefont {Jiang}},\ }\bibfield  {title} {\bibinfo {title} {Localization and mobility edges in non-hermitian disorder-free lattices},\ }\href {https://arxiv.org/abs/2306.03807} {\bibfield  {journal} {\bibinfo  {journal} {arXiv:2306.03807}\ } (\bibinfo {year} {2023}{\natexlab{b}})}\BibitemShut {NoStop}%
\bibitem [{\citenamefont {Gandhi}\ and\ \citenamefont {Bandyopadhyay}(2023)}]{gandhi2023topological}%
  \BibitemOpen
  \bibfield  {author} {\bibinfo {author} {\bibfnamefont {S.}~\bibnamefont {Gandhi}}\ and\ \bibinfo {author} {\bibfnamefont {J.~N.}\ \bibnamefont {Bandyopadhyay}},\ }\bibfield  {title} {\bibinfo {title} {Topological triple phase transition in non-hermitian quasicrystals with complex asymmetric hopping},\ }\href {https://doi.org/10.1103/PhysRevB.108.014204} {\bibfield  {journal} {\bibinfo  {journal} {Phys. Rev. B}\ }\textbf {\bibinfo {volume} {108}},\ \bibinfo {pages} {014204} (\bibinfo {year} {2023})}\BibitemShut {NoStop}%
\bibitem [{\citenamefont {Zhu}\ \emph {et~al.}(2023)\citenamefont {Zhu}, \citenamefont {Lang}, \citenamefont {Wang}, \citenamefont {Wang},\ and\ \citenamefont {Chong}}]{zhu2023topological}%
  \BibitemOpen
  \bibfield  {author} {\bibinfo {author} {\bibfnamefont {B.}~\bibnamefont {Zhu}}, \bibinfo {author} {\bibfnamefont {L.-J.}\ \bibnamefont {Lang}}, \bibinfo {author} {\bibfnamefont {Q.}~\bibnamefont {Wang}}, \bibinfo {author} {\bibfnamefont {Q.~J.}\ \bibnamefont {Wang}},\ and\ \bibinfo {author} {\bibfnamefont {Y.~D.}\ \bibnamefont {Chong}},\ }\bibfield  {title} {\bibinfo {title} {Topological transitions with an imaginary {Aubry-Andr\'e-Harper} potential},\ }\href {https://doi.org/10.1103/PhysRevResearch.5.023044} {\bibfield  {journal} {\bibinfo  {journal} {Phys. Rev. Res.}\ }\textbf {\bibinfo {volume} {5}},\ \bibinfo {pages} {023044} (\bibinfo {year} {2023})}\BibitemShut {NoStop}%
\bibitem [{\citenamefont {Jiang}\ \emph {et~al.}(2024)\citenamefont {Jiang}, \citenamefont {Zeng}, \citenamefont {Hu},\ and\ \citenamefont {Liu}}]{jiang2024exact}%
  \BibitemOpen
  \bibfield  {author} {\bibinfo {author} {\bibfnamefont {X.-P.}\ \bibnamefont {Jiang}}, \bibinfo {author} {\bibfnamefont {W.}~\bibnamefont {Zeng}}, \bibinfo {author} {\bibfnamefont {Y.}~\bibnamefont {Hu}},\ and\ \bibinfo {author} {\bibfnamefont {P.}~\bibnamefont {Liu}},\ }\bibfield  {title} {\bibinfo {title} {Exact non-hermitian mobility edges and robust flat bands in two-dimensional lieb lattices with imaginary quasiperiodic potentials},\ }\href {https://iopscience.iop.org/article/10.1088/1367-2630/ad6bb9/meta} {\bibfield  {journal} {\bibinfo  {journal} {New J. Phys.}\ } (\bibinfo {year} {2024})}\BibitemShut {NoStop}%
\bibitem [{\citenamefont {Acharya}\ and\ \citenamefont {Datta}(2024)}]{acharya2024localization}%
  \BibitemOpen
  \bibfield  {author} {\bibinfo {author} {\bibfnamefont {A.~P.}\ \bibnamefont {Acharya}}\ and\ \bibinfo {author} {\bibfnamefont {S.}~\bibnamefont {Datta}},\ }\bibfield  {title} {\bibinfo {title} {Localization transitions in a non-hermitian quasiperiodic lattice},\ }\href {https://doi.org/10.1103/PhysRevB.109.024203} {\bibfield  {journal} {\bibinfo  {journal} {Phys. Rev. B}\ }\textbf {\bibinfo {volume} {109}},\ \bibinfo {pages} {024203} (\bibinfo {year} {2024})}\BibitemShut {NoStop}%
\bibitem [{\citenamefont {Padhan}\ \emph {et~al.}(2024)\citenamefont {Padhan}, \citenamefont {Padhi},\ and\ \citenamefont {Mishra}}]{padhan2024complete}%
  \BibitemOpen
  \bibfield  {author} {\bibinfo {author} {\bibfnamefont {A.}~\bibnamefont {Padhan}}, \bibinfo {author} {\bibfnamefont {S.~R.}\ \bibnamefont {Padhi}},\ and\ \bibinfo {author} {\bibfnamefont {T.}~\bibnamefont {Mishra}},\ }\bibfield  {title} {\bibinfo {title} {Complete delocalization and reentrant topological transition in a non-hermitian quasiperiodic lattice},\ }\href {https://doi.org/10.1103/PhysRevB.109.L020203} {\bibfield  {journal} {\bibinfo  {journal} {Phys. Rev. B}\ }\textbf {\bibinfo {volume} {109}},\ \bibinfo {pages} {L020203} (\bibinfo {year} {2024})}\BibitemShut {NoStop}%
\bibitem [{\citenamefont {Wang}\ \emph {et~al.}(2024{\natexlab{a}})\citenamefont {Wang}, \citenamefont {Wang},\ and\ \citenamefont {Chen}}]{wang2024non}%
  \BibitemOpen
  \bibfield  {author} {\bibinfo {author} {\bibfnamefont {L.}~\bibnamefont {Wang}}, \bibinfo {author} {\bibfnamefont {Z.}~\bibnamefont {Wang}},\ and\ \bibinfo {author} {\bibfnamefont {S.}~\bibnamefont {Chen}},\ }\bibfield  {title} {\bibinfo {title} {Non-{Hermitian} butterfly spectra in a family of quasiperiodic lattices},\ }\href {https://doi.org/10.1103/PhysRevB.110.L060201} {\bibfield  {journal} {\bibinfo  {journal} {Phys. Rev. B}\ }\textbf {\bibinfo {volume} {110}},\ \bibinfo {pages} {L060201} (\bibinfo {year} {2024}{\natexlab{a}})}\BibitemShut {NoStop}%
\bibitem [{\citenamefont {Li}\ and\ \citenamefont {Li}(2024)}]{li2024ring}%
  \BibitemOpen
  \bibfield  {author} {\bibinfo {author} {\bibfnamefont {S.-Z.}\ \bibnamefont {Li}}\ and\ \bibinfo {author} {\bibfnamefont {Z.}~\bibnamefont {Li}},\ }\bibfield  {title} {\bibinfo {title} {Ring structure in the complex plane: A fingerprint of a non-hermitian mobility edge},\ }\href {https://doi.org/10.1103/PhysRevB.110.L041102} {\bibfield  {journal} {\bibinfo  {journal} {Phys. Rev. B}\ }\textbf {\bibinfo {volume} {110}},\ \bibinfo {pages} {L041102} (\bibinfo {year} {2024})}\BibitemShut {NoStop}%
\bibitem [{\citenamefont {Wang}\ \emph {et~al.}(2024{\natexlab{b}})\citenamefont {Wang}, \citenamefont {Liu}, \citenamefont {Wang},\ and\ \citenamefont {Chen}}]{wang2024exact}%
  \BibitemOpen
  \bibfield  {author} {\bibinfo {author} {\bibfnamefont {L.}~\bibnamefont {Wang}}, \bibinfo {author} {\bibfnamefont {J.}~\bibnamefont {Liu}}, \bibinfo {author} {\bibfnamefont {Z.}~\bibnamefont {Wang}},\ and\ \bibinfo {author} {\bibfnamefont {S.}~\bibnamefont {Chen}},\ }\bibfield  {title} {\bibinfo {title} {Exact complex mobility edges and flagellate spectra for non-hermitian quasicrystals with exponential hoppings},\ }\href {https://arxiv.org/abs/2406.10769} {\bibfield  {journal} {\bibinfo  {journal} {arXiv:2406.10769}\ } (\bibinfo {year} {2024}{\natexlab{b}})}\BibitemShut {NoStop}%
\bibitem [{\citenamefont {Longhi}(2021)}]{longhi2021non}%
  \BibitemOpen
  \bibfield  {author} {\bibinfo {author} {\bibfnamefont {S.}~\bibnamefont {Longhi}},\ }\bibfield  {title} {\bibinfo {title} {Non-{Hermitian} skin effect beyond the tight-binding models},\ }\href {https://doi.org/10.1103/PhysRevB.104.125109} {\bibfield  {journal} {\bibinfo  {journal} {Phys. Rev. B}\ }\textbf {\bibinfo {volume} {104}},\ \bibinfo {pages} {125109} (\bibinfo {year} {2021})}\BibitemShut {NoStop}%
\bibitem [{\citenamefont {Mochizuki}\ and\ \citenamefont {Ozawa}(2022)}]{mochizuki2022band}%
  \BibitemOpen
  \bibfield  {author} {\bibinfo {author} {\bibfnamefont {K.}~\bibnamefont {Mochizuki}}\ and\ \bibinfo {author} {\bibfnamefont {T.}~\bibnamefont {Ozawa}},\ }\bibfield  {title} {\bibinfo {title} {Band structures under non-{Hermitian} periodic potentials: Connecting nearly-free and bi-orthogonal tight-binding models},\ }\href {https://doi.org/10.1103/PhysRevB.105.174108} {\bibfield  {journal} {\bibinfo  {journal} {Phys. Rev. B}\ }\textbf {\bibinfo {volume} {105}},\ \bibinfo {pages} {174108} (\bibinfo {year} {2022})}\BibitemShut {NoStop}%
\bibitem [{\citenamefont {Yokomizo}\ \emph {et~al.}(2022)\citenamefont {Yokomizo}, \citenamefont {Yoda},\ and\ \citenamefont {Murakami}}]{yokomizo2022non}%
  \BibitemOpen
  \bibfield  {author} {\bibinfo {author} {\bibfnamefont {K.}~\bibnamefont {Yokomizo}}, \bibinfo {author} {\bibfnamefont {T.}~\bibnamefont {Yoda}},\ and\ \bibinfo {author} {\bibfnamefont {S.}~\bibnamefont {Murakami}},\ }\bibfield  {title} {\bibinfo {title} {Non-{Hermitian} waves in a continuous periodic model and application to photonic crystals},\ }\href {https://doi.org/10.1103/PhysRevResearch.4.023089} {\bibfield  {journal} {\bibinfo  {journal} {Phys. Rev. Res.}\ }\textbf {\bibinfo {volume} {4}},\ \bibinfo {pages} {023089} (\bibinfo {year} {2022})}\BibitemShut {NoStop}%
\bibitem [{\citenamefont {Yuce}\ and\ \citenamefont {Ramezani}(2022{\natexlab{b}})}]{yuce2022non}%
  \BibitemOpen
  \bibfield  {author} {\bibinfo {author} {\bibfnamefont {C.}~\bibnamefont {Yuce}}\ and\ \bibinfo {author} {\bibfnamefont {H.}~\bibnamefont {Ramezani}},\ }\bibfield  {title} {\bibinfo {title} {Non-hermitian skin effect in two dimensional continuous systems},\ }\href {https://doi.org/10.1088/1402-4896/aca43b} {\bibfield  {journal} {\bibinfo  {journal} {Phys. Scr.}\ }\textbf {\bibinfo {volume} {98}},\ \bibinfo {pages} {015005} (\bibinfo {year} {2022}{\natexlab{b}})}\BibitemShut {NoStop}%
\bibitem [{\citenamefont {Zeng}\ and\ \citenamefont {Lü}(2023)}]{zeng2023gaussian}%
  \BibitemOpen
  \bibfield  {author} {\bibinfo {author} {\bibfnamefont {Q.-B.}\ \bibnamefont {Zeng}}\ and\ \bibinfo {author} {\bibfnamefont {R.}~\bibnamefont {Lü}},\ }\bibfield  {title} {\bibinfo {title} {Gaussian eigenstate pinning in non-{Hermitian} quantum mechanics},\ }\href {https://doi.org/10.1103/PhysRevA.107.062221} {\bibfield  {journal} {\bibinfo  {journal} {Phys. Rev. A}\ }\textbf {\bibinfo {volume} {107}},\ \bibinfo {pages} {062221} (\bibinfo {year} {2023})}\BibitemShut {NoStop}%
\bibitem [{\citenamefont {Hu}\ \emph {et~al.}(2023)\citenamefont {Hu}, \citenamefont {Huang}, \citenamefont {Xue},\ and\ \citenamefont {Wang}}]{hu2023non}%
  \BibitemOpen
  \bibfield  {author} {\bibinfo {author} {\bibfnamefont {Y.-M.}\ \bibnamefont {Hu}}, \bibinfo {author} {\bibfnamefont {Y.-Q.}\ \bibnamefont {Huang}}, \bibinfo {author} {\bibfnamefont {W.-T.}\ \bibnamefont {Xue}},\ and\ \bibinfo {author} {\bibfnamefont {Z.}~\bibnamefont {Wang}},\ }\bibfield  {title} {\bibinfo {title} {Non-{Bloch} band theory for non-{Hermitian} continuum systems},\ }\href {https://arxiv.org/abs/2310.08572} {\bibfield  {journal} {\bibinfo  {journal} {arXiv:2310.08572}\ } (\bibinfo {year} {2023})}\BibitemShut {NoStop}%
\bibitem [{\citenamefont {Yao}\ \emph {et~al.}(2019)\citenamefont {Yao}, \citenamefont {Khoudli}, \citenamefont {Bresque},\ and\ \citenamefont {Sanchez-Palencia}}]{khoudli2019critical}%
  \BibitemOpen
  \bibfield  {author} {\bibinfo {author} {\bibfnamefont {H.}~\bibnamefont {Yao}}, \bibinfo {author} {\bibfnamefont {H.}~\bibnamefont {Khoudli}}, \bibinfo {author} {\bibfnamefont {L.}~\bibnamefont {Bresque}},\ and\ \bibinfo {author} {\bibfnamefont {L.}~\bibnamefont {Sanchez-Palencia}},\ }\bibfield  {title} {\bibinfo {title} {Critical behavior and fractality in shallow one-dimensional quasiperiodic potentials},\ }\href {https://doi.org/10.1103/PhysRevLett.123.070405} {\bibfield  {journal} {\bibinfo  {journal} {Phys. Rev. Lett.}\ }\textbf {\bibinfo {volume} {123}},\ \bibinfo {pages} {070405} (\bibinfo {year} {2019})}\BibitemShut {NoStop}%
\bibitem [{\citenamefont {Deng}\ \emph {et~al.}(2019)\citenamefont {Deng}, \citenamefont {Ray}, \citenamefont {Sinha}, \citenamefont {Shlyapnikov},\ and\ \citenamefont {Santos}}]{deng2019one}%
  \BibitemOpen
  \bibfield  {author} {\bibinfo {author} {\bibfnamefont {X.}~\bibnamefont {Deng}}, \bibinfo {author} {\bibfnamefont {S.}~\bibnamefont {Ray}}, \bibinfo {author} {\bibfnamefont {S.}~\bibnamefont {Sinha}}, \bibinfo {author} {\bibfnamefont {G.~V.}\ \bibnamefont {Shlyapnikov}},\ and\ \bibinfo {author} {\bibfnamefont {L.}~\bibnamefont {Santos}},\ }\bibfield  {title} {\bibinfo {title} {One-dimensional quasicrystals with power-law hopping},\ }\href {https://doi.org/10.1103/PhysRevLett.123.025301} {\bibfield  {journal} {\bibinfo  {journal} {Phys. Rev. Lett.}\ }\textbf {\bibinfo {volume} {123}},\ \bibinfo {pages} {025301} (\bibinfo {year} {2019})}\BibitemShut {NoStop}%
\end{thebibliography}%
\end{document}